\documentclass[10pt,journal,twoside]{IEEEtran}

\usepackage[utf8]{inputenc}
\usepackage[cmex10]{amsmath}
\usepackage{graphicx}
\usepackage{amssymb}
\usepackage{amsthm}
\usepackage{subfigure}
\usepackage{cite}
\usepackage{color}
\usepackage{bm}
\usepackage{dblfloatfix}
\usepackage[mathlines]{lineno}

\usepackage[author={M. Kokshoorn}]{pdfcomment}
\usepackage{xcolor}

\theoremstyle{remark}

\usepackage[ruled,norelsize]{algorithm2e}
\makeatletter
\newcommand{\removelatexerror}{\let\@latex@error\@gobble}
\makeatother
\usepackage{algorithmicx}

\def \edit {}

\begin{document}


\newcommand*{\TitleFont}{%
      \fontsize{20}{20}%
      \selectfont}

\title{\TitleFont{Beam-On-Graph: Simultaneous Channel Estimation for mmWave MIMO Systems with Multiple Users}}

\author{
    {Matthew Kokshoorn, He Chen, Yonghui Li, and Branka Vucetic}
\thanks{This work was supported in part by the Australian Research Council under grant  DP150104019. The work of Y. Li was also supported by NSFC under grants 61531006 and 61772233.}
\thanks{The material in this paper has been presented in part at the IEEE International Conference on Communications, Paris, France, May 2017 \cite{kokshoorn2016fountain}.}
\thanks{
M. Kokshoorn, H. Chen, Y. Li, and B. Vucetic are with School of Electrical and Information Engineering, The University of Sydney, Sydney, NSW 2006, Australia (email: \{matthew.kokshoorn, he.chen, yonghui.li, branka.vucetic\}@sydney.edu.au).}
}

\maketitle

\begin{abstract}
This paper is concerned with the channel estimation problem in multi-user millimeter wave (mmWave) wireless systems with large antenna arrays. We develop a novel simultaneous-estimation with iterative fountain training (SWIFT) framework, in which multiple users estimate their channels at the same time and the required number of channel measurements is adapted to various channel conditions of different users. To achieve this, we represent the beam direction estimation process by a graph, referred to as the beam-on-graph, and associate the channel estimation process with a code-on-graph decoding problem. Specifically, the base station (BS) and each user measure the channel with a series of random combinations of transmit/receive beamforming vectors until the channel estimate converges. As the proposed SWIFT does not adapt the BS's beams to any single user, we are able to estimate all user channels simultaneously. Simulation results show that SWIFT can significantly outperform the existing random beamforming-based approaches, which use a predetermined number of measurements, over a wide range of signal-to-noise ratios and channel coherence time. Furthermore, by utilizing the users' order in terms of completing their channel estimation, our SWIFT framework can infer the sequence of users' channel quality and perform effective user scheduling to achieve superior performance.
\end{abstract}


\section{Introduction}

Due to the increasing congestion in the microwave spectrum, alternative frequencies are now being considered for 5G cellular systems \cite{rappaport2013millimeter,dehos2014millimeter,roh2014millimeter}. \edit{More specifically, millimeter wave (mmWave) frequencies, ranging from 30GHz to 300GHz, have recently attracted significant attention due to the wide expanse of underutilized bandwidth \cite{wong2017key,ge20165g,zeng2016millimeter}}. One fundamental issue of mmWave communications stems from the large free space propagation loss experienced by signals in the high frequency range \cite{rappaportMeasure,heath2016overview}. Supplementing this issue, penetration and reflection losses are also much more significant than those at microwave frequencies. As such, the mmWave channel is relatively sparse in the spatial domain, with only a limited number of propagation path directions suitable for conveying information. Overcoming these challenges is now more than ever essential to best utilize the mmWave spectrum, e.g., the 14GHz of the unlicensed spectrum and the 3.85GHz of licensed spectrum recently made available by the FCC in the United States \cite{FFCQual}.

The most widely accepted means to overcome and even exploit the inherent mmWave weaknesses, is to implement large antenna arrays so that narrow beams with high beamforming gains can be generated to overcome the severe signal losses \cite{bj2016massive}. Thanks to the small wavelength of the mmWave band, these large arrays can still maintain a small form factor. The general idea of mmWave communications is then to steer these narrow beams in the direction of the available propagation paths, effectively ``bouncing'' information-bearing signals off buildings and various other scatterers. As a result, in mmWave communication systems, the propagation paths are normally estimated through directly finding the beam-steering directions of each of path \cite{rheath}.

\begin{figure}[!t]
\centering
\includegraphics[width=3.3in,trim={0.5cm 10cm 1.0cm 7.0cm},clip]{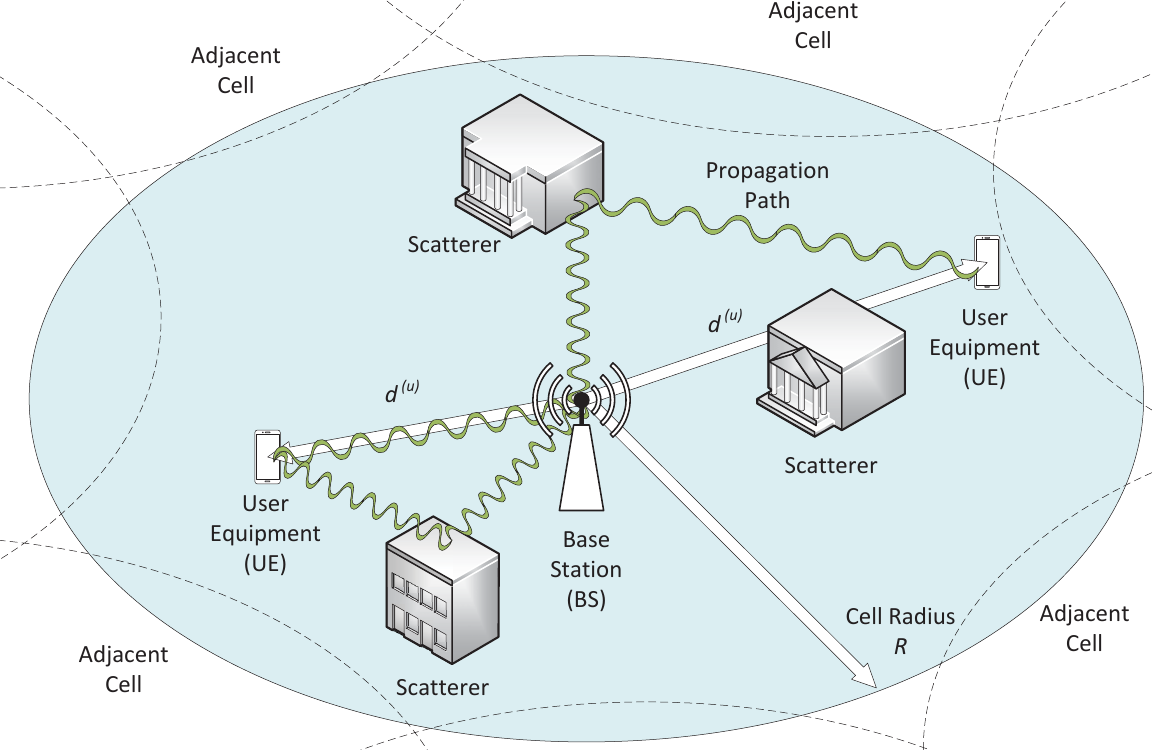}
\caption{An example of multi-user mmWave cellular systems.}
\label{deployment_scenario}
\end{figure}

In conventional low-bandwidth microwave MIMO systems, fully digital hardware with an RF chain associated with each antenna, is able to implement digital control/sampling of the phase and amplitude of the baseband signal from each antenna. However, in mmWave communication systems with large antenna arrays, equipping every antenna with an individual radio frequency (RF) chain along with high frequency analog-to-digital converter (ADC) and digital-to-analog converter (DAC) would incur high hardware cost, complexity and power consumption, particularly in the context of consumer electronics. Fortunately, due to the limited number of propagation paths in the mmWave links, it has been widely recognized that a fully digital system (i.e., dedicated RF chain for each antenna) is not necessary \cite{rheath}. Instead, networks of phase shifters can be used to adjust the phase of the transmitted or received signal on each antenna to realize transmit or receive beamforming. The input/output of each group of phase shifters is then tied to a common RF chain. Although this setup reduces the hardware cost and complexity, it restricts the system to the use of RF beamforming and can thus only send or receive signals with a single beamforming vector (i.e., a set of phase shifts) for each RF chain. To further simplify the hardware requirements, the phase shifts are often limited to a quantized set of values\cite{alkhateeb2015limited,rheath}, resulting in only a finite number of possible beamforming directions. Some recent work has even considered reducing hardware complexity further by using one-bit ADCs \cite{Jianhua}.



Adhering to these constraints and leveraging the sparse characteristic of mmWave geometric channels, previous work has focused on “divide and conquer" type multi-stage algorithms to estimate mmWave channels \cite{rheath,Kokshoorn,7579573,kokshoorn2016race}. These algorithms are essentially path finding schemes, which divide the process of finding each propagation path into multiple stages. In each subsequent stage, as the user feeds back information to the base station (BS), the estimated angular range is refined so that narrower beam patterns can be used in each following set of channel measurements. These multi-stage approaches have been shown to work efficiently for point-to-point mmWave communications \cite{rheath,Kokshoorn,7579573,kokshoorn2016race}. However, by adapting the BS beam patterns to a specific user, these approaches are inherently limited to estimating only a small number of users in each channel estimation process. As a result, for multi-user scenarios, these types of approaches may no longer be efficient as they could require a training overhead that scales linearly with the number of users \cite{alkhateeby2015compressed}.

Different from these multi-stage adaptive channel estimation algorithms, multi-user beamforming-based approaches are able to carry out simultaneous channel estimation for multiple channels. \edit{In \cite{zhao2017multi}, the authors proposed a frequency tone-based estimation and ZF precoding strategy for multi-user channels}. Random compressed sensing-based channel estimation approaches using random beam-directions have been explored in \cite{mendez2015channel,ramasamy2012compressive,ramasamy2012compressive2,berraki2014application,alkhateeby2015compressed}. These random beamforming-based channel estimation approaches generally perform a predetermined number of random channel measurements before the channel estimation decision is made.  However, selecting a fixed number of channel measurements may not work well for all users and channel realizations, and can thus lead to an inferior system performance. For example, in a channel realization resulting in high signal-to-noise ratio (SNR), the channel estimation may not require as many measurements as they would at low SNR. This phenomenon for a multi-user scenario has been discovered in \cite{alkhateeby2015compressed}, wherein different numbers of measurements are required for users with different lengths of coherence time and SNRs. However, in reality, multi-user scenarios such as that in Fig. \ref{deployment_scenario} can have users with distinct SNRs as they may have different channel characteristics to the BS. As such, it may not be feasible to achieve an optimal channel training duration that is commonly suitable for all users by adopting a fixed number of measurements. \edit{Adding to this challenge, even in low mobility scenarios (e.g, walking less than 1.5m/s), channel measurements for carrier frequencies at 60 GHz have been shown to exhibit channel coherence times less than 1ms \cite{smulders2009statistical}. At slightly faster speeds of 3m/s, 30 GHz measurements have also shown coherence times of 1.5ms \cite{marinier1998temporal}. With proposed OFDM data symbol durations on the order of few microseconds \cite{baldemair2015ultra,khan2012millimeter}, the mmWave channel may only have coherency on the order of hundreds of symbols \cite{park2007short}.}

In digital communication systems, the problem of adapting the mmWave channel training duration is analogous to adapting the transmission rate of communication systems to real-time unknown channel conditions. That is, we seek to adapt the number of channel estimation measurements without any prior knowledge of various channel realizations of multiple users. The conventional rate adaptation problem has led to the development of a powerful rateless coding family known as fountain codes. Inspired by the principle of analog fountain codes (AFC) \cite{shirvanimoghaddam2013near}, in this paper we develop a novel Simultaneous-estimation With Iterative Fountain Training (SWIFT) framework for the channel estimation of multi-user mmWave MIMO systems. In SWIFT, the training duration required for estimating the multi-user channels is adaptively increased until a predetermined stopping criterion has been met at each user.

In this paper, we represent the random beamforming process by a graph, called beam-on-graph and match the beam-on-graph to a code-on-graph. Specifically, we propose a Fountain code-like channel estimation approach, in which the BS keeps transmitting pilot signals in random beam-directions for an indefinite period, essentially ``encoding'' random pieces of the virtual channel information in each measurement. At the same time, all users within the BS coverage keep “listening” for these pilot signals by receiving them with random beam-directions. After each measurement, each user estimates its channel based on the pilot signals it has collected, and compares it to the previous estimate. If the estimate is similar to the previous estimate (i.e., the estimate has converged), the user regards its channel estimation procedure as complete. The user then feeds back the indices of the BS beamforming vectors to be adopted for its data communication. We summarize the main contributions of this paper as follows:

\begin{itemize}
\item{
We propose a novel SWIFT algorithm to realize simultaneous multi-user channel estimation in mmWave systems, where the average number of channel measurements is adapted to different channel conditions of multiple users. We measure the mmWave channels using a variable number of beam patterns until the channel estimate of each user converges. We develop a framework to optimize the random beamforming process by matching the beam-on-graph to a code-on-graph. We formulate the estimation of each channel as a compressed sensing problem and implement a generalized approximate message passing (GAMP) \cite{rangan2011generalized} algorithm to recover the sparse virtual channel information.
}
\item{Although the beamforming directions at the BS cannot be adaptive to a particular user to ensure the simultaneous channel estimation of multiple users, this does not restrict the adaptation of receiving beam-directions at the user side. Motivated by this, we propose two user-side beamforming adaptation schemes to further improve the estimation performance.
}
\item{
We compare the proposed algorithm with the existing random beamforming-based approaches with a predefined number of measurements. Simulation results show that the proposed SWIFT algorithm can outperform random beamforming-based approaches, over a range of SNRs and coherence time. Furthermore, by utilizing the users' order in terms of completing their channel estimation, our SWIFT framework can infer the users' channel quality and perform effective user scheduling to achieve superior rate performance, especially for resource-constrained scenarios where only a limited number of users can be served.
}
\end{itemize}

{\textbf{Notations}} : We use letter $\bm{A}$ to denote a matrix, $\bm{a}$ to denote a vector, ${a}$ to denote a scalar, and $\mathcal{A}$ denotes a set. $|a|$ is the absolute value of $a$, $||\bm{A}||_2$ is the 2-norm of $\bm{A}$ and $\text{det}(\bm{A})$ is the determinant of $\bm{A}$. $\bm{A}^T$, $\bm{A}^H$ and $\bm{A}^*$ are the transpose, conjugate transpose and conjugate of $\bm{A}$, respectively. For a square matrix $\bm{A}$, $\bm{A}^{-1}$ represents its inverse. $\bm{I}_N$ is the $N\times N$ identity matrix and $\lceil \cdot \rceil$ denotes the ceiling function. $\mathcal{C}\mathcal{N}(\bm{m},\bm{R})$ is a complex Gaussian random vector with mean $\bm{m}$ and covariance matrix $\bm{R}$, and $\text{E}[\bm{a}]$ and $\text{Cov}[\bm{a}]$ denote the expected value and covariance of ${\bm{a}}$, respectively.

\section{System Model}
Consider a multi-user mmWave MIMO system comprising of a BS with $N_{\!B\!S}$ antennas and $U$ sets of user equipment (UE), each with $N_{\!U\!E}$ antennas. We consider that the BS and UE are equipped with a limited number of RF chains, denoted by $R_{\!B\!S}$ and $R_{\!U\!E}$, respectively. To estimate the downlink channel matrix, the BS broadcasts a sequence of beamformed pilot signals to all UEs at the same time. Denote by $\bm{f}_i$ the $N_{\!B\!S} \times 1$ transmitting beamforming vector adopted by the $i$th RF chain at the BS. Similarly, denote by $\bm{w}_j^{(u)}$, the $N_{\!U\!E} \times 1$ receiving beamforming vector adopted by the $j$th RF chain of the $u$th user.

\begin{figure}[!t]
\centering
\includegraphics[width=3.3in,trim={0.5cm 7.4cm 2.0cm 2.1cm},clip]{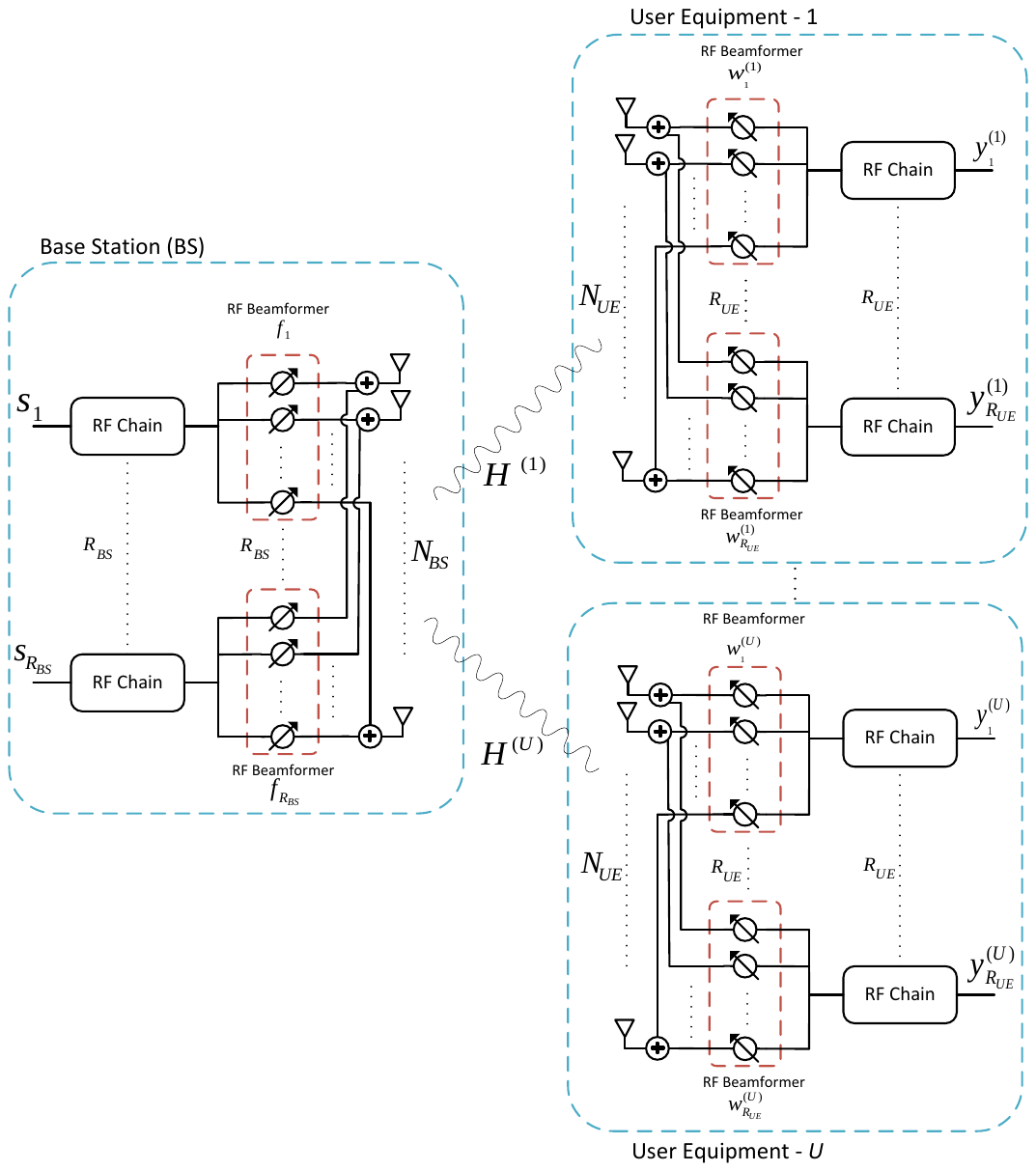}
\caption{System model of the considered multi-user mmWave MIMO system.}
\label{system_model}
\end{figure}

Here, we consider the beamforming vectors, at each link end, to be limited to networks of RF phase shifters \cite{rheath} as shown in Fig. \ref{system_model}. As such, all elements of $\bm{f}_i$ and $\bm{w}_i^{(u)}$ have constant modulus and unit norm such that $||\bm{f}_i||=1, \forall$ $i =1,\cdots ,R_{\!B\!S}$, and $||\bm{w}_j^{(u)}||=1, \forall$ $j =1,\cdots ,R_{\!U\!E}, u =1,\cdots ,U$. We further assume that due to hardware constraints, each phase shifter (i.e., the entries of $\bm{f}_i$ and $\bm{w}_j^{(u)}$ ) is digitally controlled and can only use quantized values from a predetermined set given by
\begin{align} \label{BS_set}
\left\{ \frac{1}{\sqrt{N}}\text{exp}(j q_k) \right\}, \forall k=1,\cdots,N,
\end{align}
\edit{where $N\in\{N_{\!B\!S},N_{\!U\!E}\}$ is the number of antennas in the array. The set, $\{q_k=\pi-2\pi(k-1)/N \forall k \}$ describes the $N$ quantized phase shift angles which, by convention, are defined to start at $\pi$ and are uniformly spaced clockwise around the unit circle.} That is, each BS (UE) phase shifter can only use one of $N_{\!B\!S}$ ($N_{\!U\!E}$) uniformly spaced phase shifts, respectively, and can therefore be digitally controlled by $\lceil\text{log}_2N\rceil$ bits.
%

Let $\bm{F}=[\bm{f}_1,\bm{f}_2,\cdots,\bm{f}_{R_{\!B\!S}}]$ denote the $N_{\!B\!S}\times R_{\!B\!S}$ BS beamforming matrix, with columns representing the $R_{\!B\!S}$ RF beamforming vectors. The corresponding $N_{\!B\!S}\times 1$ BS transmit signal can be represented as
\begin{equation} \label{x}
\bm{x} = \sqrt{\frac{P}{R_{\!B\!S}}} \bm{F} \bm{s},
\end{equation}
where $P$ is the total transmit power of the BS and $\boldsymbol{s}$ is the ${R}_{\!B\!S}\times 1$ vector of transmit pilot symbols corresponding to $R_{BS}$ numbers of beamforming vectors with $E[\bm{s}\bm{s}^H]=\bm{I}_{R_{\!B\!S}}$. We adopt a widely-used block-fading channel model such that the signal observed by the $u$th user can be expressed as \cite{alkhateeby2015compressed}
\begin{align} \label{r}
\bm{r}^{(u)} &= \bm{H}^{(u)}\bm{x} +\bm{q}^{(u)} = \sqrt{\frac{P}{R_{\!B\!S}}} \bm{H}^{(u)}\bm{F} \bm{s} + \bm{q}^{(u)},
\end{align}
where $\bm{H}^{(u)}$ denotes the $N_{\!U\!E}\times N_{\!B\!S}$ MIMO channel matrix between the BS and the $u$th user, and $\bm{q}^{(u)}$ is an $N_{\!U\!E} \times 1$ complex additive white Gaussian noise (AWGN) vector for the $u$th user following distribution $\mathcal{C}\mathcal{N}(0, N_0 \bm{I}_{N_{\!U\!E}})$.

Each user processes the received pilot signals with each of the $R_{\!U\!E}$ RF chains. By denoting $\bm{W}^{(u)}=[\bm{w}_1^{(u)},\bm{w}_2^{(u)},\cdots,\bm{w}_{R_{\!U\!E}}^{(u)}]$ as the $N_{\!U\!E}\times R_{\!U\!E}$ combining matrix at the $u$th user, we express the $R_{\!U\!E} \times 1$ vector of the $u$th user's received signals as
\begin{align} \label{y}
\bm{y}^{(u)} &= (\bm{W}^{(u)})^H \bm{H}^{(u)}\bm{x} +\boldsymbol{n}^{(u)}
\end{align}
where, since $||\bm{w}_j^{(u)}||_2 = 1$, $\forall$ $j$, the vector $\boldsymbol{n}^{(u)}=(\bm{W}^{(u)})^H\bm{q}^{(u)} $ follows the distribution as that of ${\bm{q}^{(u)}}$, i.e., $\bm{n}^{(u)} \sim \mathcal{C}\mathcal{N}(0, N_0 (\bm{W}^{(u)})^H  \bm{W}^{(u)})$. 

We follow \cite{Sayeed_max} and adopt a two-dimensional (2D) sparse geometric-based channel model. Specifically, we consider that there are $L^{(u)}$ paths between the BS and the $u$th user, with the $u$th user's $l$th path having AOD, $\phi_l^{(u)}$, and AOA, $\theta_l^{(u)}$ with $l=1,...,L^{(u)}$. We further consider these AOD/AOA to be uniformly distributed on the range $[0,2\pi)$. Then the corresponding channel matrix can be expressed in terms of the physical propagation path parameters as
%
\begin{equation} \label{H}
\bm{H}^{(u)} = \sqrt{N_{\!B\!S}N_{\!U\!E}}\sum\limits_{l=1}^{L^{(u)}}\alpha_l^{(u)}   \bm{a}_{\!U\!E}(\theta_l^{(u)}) (\bm{a}_{\!B\!S}(\phi_l^{(u)}))^H
\end{equation}
%

\noindent
where $\alpha_l^{(u)}\sim\mathcal{C}\mathcal{N}(0,\sigma_R^{(u)})$ is the channel fading coefficient of the $l$th propagation path of the $u$th user, and $\bm{a}_{\!B\!S}(\theta_l^{(u)})$ and $\bm{a}_{\!U\!E}(\phi_l^{(u)})$ denote the BS and UE spatial signatures of the $l$th path, respectively. For the purpose of exploration, we consider that the BS and each UE are equipped with linear antenna arrays (ULA). However, it is worth pointing out that the developed scheme can be easily extended to other antenna structures. Using ULAs, we can define $\bm{a}_{\!B\!S}(\phi_l^{(u)})= \bm{u}(\phi_l^{(u)},N_{\!B\!S})$ and $\bm{a}_{\!U\!E}(\theta_l^{(u)}) = \bm{u}(\theta_l^{(u)},N_{\!U\!E})$, respectively, where
%
\begin{equation} \label{u}
\bm{u}(\epsilon,N) \triangleq \frac{1}{\sqrt{N}} [1,e^{j \frac{2 \pi d \text{cos}(\epsilon)}{\lambda} },\cdots,e^{j \frac{2\pi d (N-1) \text{cos}(\epsilon)}{\lambda} }]^T.
\end{equation}
\noindent
In (\ref{u}), $N\in\{N_{\!B\!S},N_{\!U\!E}\}$ is the number of antenna elements in the array, $\lambda$ denotes the signal wavelength and $d$ denotes the spacing between antenna elements. With half-wavelength spacing, the distance between antenna elements satisfies $d=\lambda/2$. Following the practical measurements from \cite{Akdeniz}, we model the number of paths $L^{(u)}$, as a Poisson random variable with the expected value $\text{E}[L^{(u)}]$. Then, the probability that there are $L$ paths between the BS and the $u$th user is given by
\begin{equation} \label{L_dist}
\text{Pr}(L^{(u)}=L) = \frac{(\text{E}[L^{(u)}])^{L}}{L!}\text{exp}(-\text{E}[L^{(u)}]).
\end{equation}

To estimate the channel information, at each link end we use beamforming vectors selected from a predetermined set of candidate beamforming vectors. We define the candidate beamfoming matrices as $\bm{F}_c$ and $\bm{W}_c$,  whose columns comprise of all candidate beamforming vectors at the BS and UE, respectively. For the ease of practical implementation, we consider the candidate beams to be the set of all possible orthogonal beamforming vectors that may later be used for data communication, subject to the quantized phase shifting constraints\footnote{Although we use the hardware limited set of beamforming vectors for ULA, the framework developed in this paper can be used to estimate the channel gains between any set of orthogonal candidate beamforming vectors for arbitrary antenna arrays.}. Following (\ref{BS_set}), this leads to $N_{\!B\!S}$ transmitting candidate beams and $N_{\!U\!E}$ receiving candidate beams. The $N_{\!U\!E} \times N_{\!B\!S}$ matrix formed by the product of the MIMO channel and these two candidate beamforming matrices is commonly referred to as the virtual channel matrix \cite{rheath} given by
\begin{equation} \label{H_v}
\bm{H}_v^{(u)}=  \frac{1}{\sqrt{N_{\!B\!S}N_{\!U\!E}}} (\bm{W}_c)^H  \bm{H}^{(u)}\bm{F}_c.
\end{equation}

\begin{figure}[!t]
\centering
\includegraphics[width=3.3in,trim={3.0cm 13.5cm 5cm 0.8cm},clip]{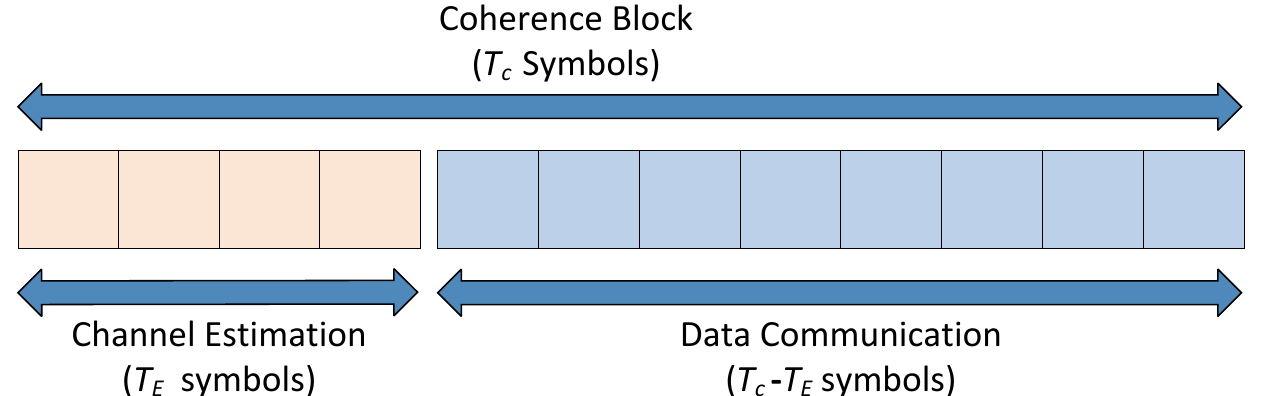}
\caption{An illustration of time allocation between channel estimation and data communication in one coherence block.}
\label{timing_model}
\end{figure}

We therefore aim to estimate this matrix so that beam pairs that result in strong channel gains can be selected out for data communication. The key challenge here is how to design a sequence of beamforming vectors in such a way that the channel parameters can be quickly and accurately estimated, leaving more time for data communication and thus achieving a higher throughput. We assume a block channel fading model with each channel realization following (\ref{H})-(\ref{L_dist}) and having coherence time of $T_c$ symbols. As coherence time is usually quite low for the mmWave frequencies (e.g., in the order of hundreds of symbols \cite{alkhateeby2015compressed}) the channel estimation time needs to be kept as short as possible to leave more time for ensuing data communication, as illustrated in Fig. \ref{timing_model}. Motivated by the fact that different users may operate in different SNR regions, in next section we develop a fountain code-inspired channel estimation algorithm for the considered multi-user mmWave system, which is able to adapt the number of channel estimation pilot symbols to various channel conditions of different users. \edit{In this paper, we consider a system with fixed spectral efficiency, array transmit power, pilot symbol duration, and baseband pilot bandwidth. As such, the primary performance parameters can be adjusted to improve the estimation performance include the total number estimation timeslots $T_E$ and the selection of candidate beams in each measurement. In practice, it would also be possible to consider a similar accuracy trade off with other domains. For example, pilot frequency division, or orthogonal pilot preambles.}

\section{The SWIFT Framework}	

In this section, we first design a set of candidate beamforming vectors to be used in our proposed channel estimation algorithm. We then formulate the channel estimation process as a compressed sensing problem and apply a sparse estimation approach to recover the virtual channel information. Finally, leveraging the introduced beam design and channel information recovery scheme, we elaborate the proposed SWIFT framework.

\subsection{Candidate Beamforming Vectors}

\begin{figure*}[!t]
\centering
\subfigure[]{\includegraphics[width=3.1in,trim={3.75cm 8.8cm 2.5cm 9.2cm},clip]{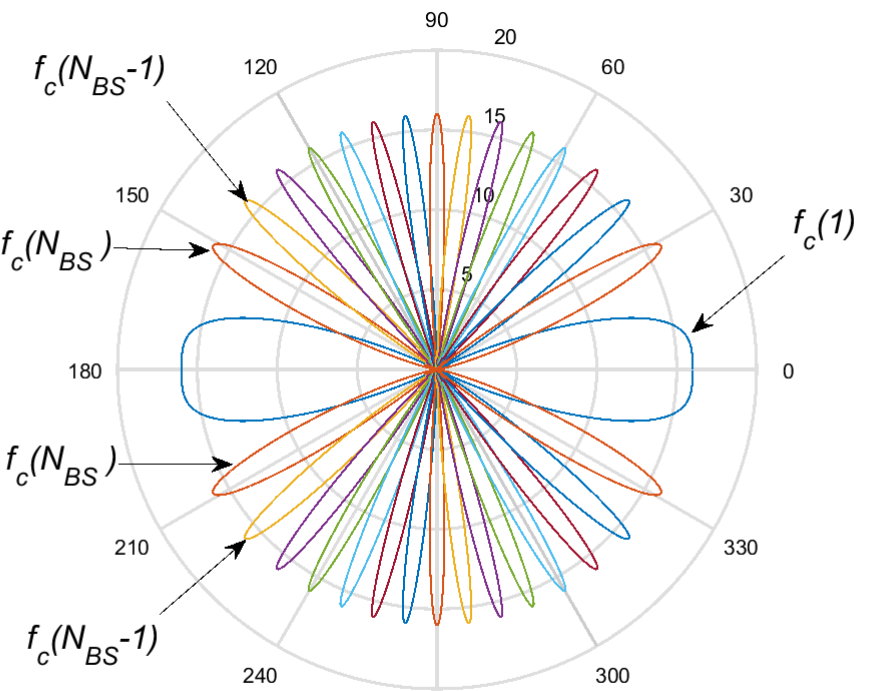}}
\subfigure[]{\includegraphics[width=3.1in,trim={3.75cm 8.8cm 2.5cm 9.2cm},clip]{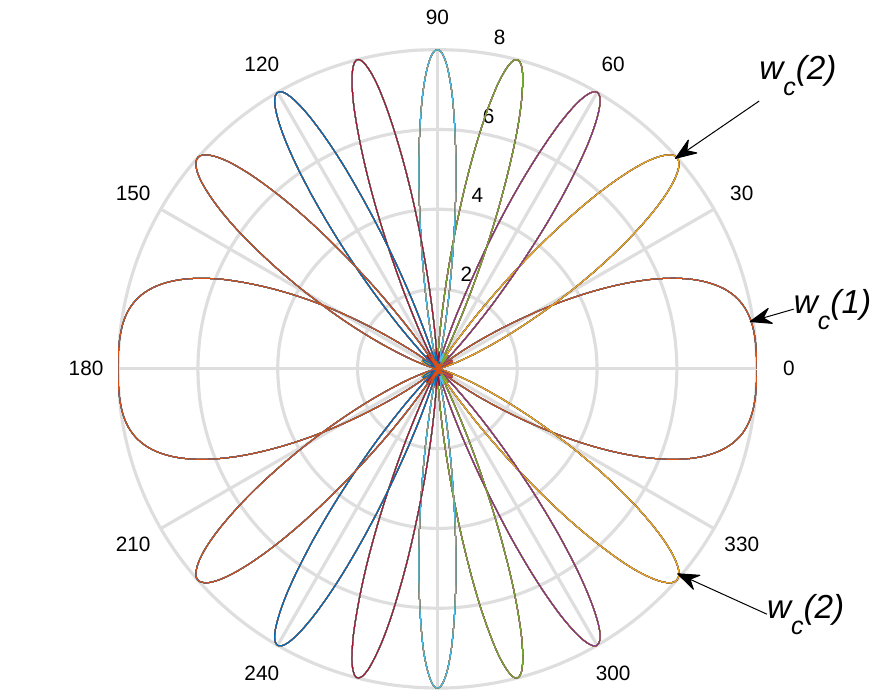}}
\caption{Example set of candidate beamforming vectors at (a) the BS with $N_{\!B\!S}=16$ and (b) the UE with $N_{\!U\!E}=8$. Due to the symmetry of ULA, beam patterns are reflected upon the range 0 to 180 degrees. In both (a) and (b) the first candidate beam can be seen at zero degrees with increasing numbered candidate beams observed in the anti-clockwise direction to 180 degrees.}
\label{candidate_beams}
\end{figure*}

We now design two sets of candidate beamforming vectors to span the full angular range using quantized phase shifters for the BS and UEs, respectively. We express the BS candidate beamforming matrix defined in (\ref{H_v}) as $\bm{F}_{c}=[\bm{f}_c(1),...,\bm{f}_c(N_{\!B\!S})]$ and the UE candidate beamforming matrix as $\bm{W}_{c}=[\bm{w}_c(1),...,\bm{w}_c(N_{\!U\!E})]$.

We define a set of candidate beam steering angles as $\bar{\epsilon}_n, \forall \; n=1,\cdots,N$ with $N\in\{ N_{\!B\!S}, N_{\!U\!E} \}$ where each corresponding beam steering vector can be expressed as $\bm{u}{({ \bar{\epsilon} }_n,N)}$ defined in (\ref{u}). \edit{Following \cite{Sayeed}, in order to satisfy the quantized phase shifter constraint, we require each beam steering vector to have values from the set $\{q_k, \forall k=1,\cdots,N\}$. Recalling that $d=\lambda/2$, by comparing the entries in (\ref{u}) with the set of quantized phase shifts, it can be seen that each steering angle $\bar{\epsilon}_n$ must satisfy the relationship
\begin{align} \label{q_eq}
 \Big\{ \bar{\epsilon}_n \Big| \frac{2 \pi d}{\lambda} cos(\bar{\epsilon}_n) \in \{ q_k \forall k \}  \Big\}
\end{align}
which leads to $\bar{\epsilon}_n = \text{cos}^{-1}(q_n/\pi)$.} Using this result, the $n$th BS candidate beamforming vector can then be described by
\begin{align} \label{f_c}
\bm{f}_c(n)=\bm{u}\Big(\text{cos}^{-1}(q_n/\pi),N_{\!B\!S}\Big), \forall n=1,\cdots,N_{\!B\!S}
\end{align}
and the $n$th UE candidate beamforming vector can be written as
\begin{align} \label{w_c}
 \bm{w}_c(n)=\bm{u}\Big(\text{cos}^{-1}(q_n/\pi),N_{\!U\!E}\Big), \forall n=1,\cdots,N_{\!U\!E}
\end{align}

As the quantized phase shifts are selected from a set of equally spaced points around the unit circle, the columns in both candidate beamforming matrices form an orthogonal set and therefore satisfy the properties $ \bm{F}_{c} \bm{F}_{c}^H=\bm{F}_{c}^H \bm{F}_{c}=\bm{I}_{N_{\!B\!S}}$ and  $\bm{W}_{c} \bm{W}_{c}^H=\bm{W}_{c}^H \bm{W}_{c}=\bm{I}_{N_{\!U\!E}}$. That is, $\bm{F}_c$ itself  and its conjugate transpose $\bm{F}_c^H$, are each equal to their own inverse. We illustrate an example set of candidate beamforming vectors for a scenario with $N_{\!B\!S}=16$ and $N_{\!U\!E}=8$ in Fig. \ref{candidate_beams}. 

\begin{figure}[!t]
\centering
\includegraphics[width=3.3in,trim={2.0cm 1.4cm 1.0cm 0.7cm},clip]{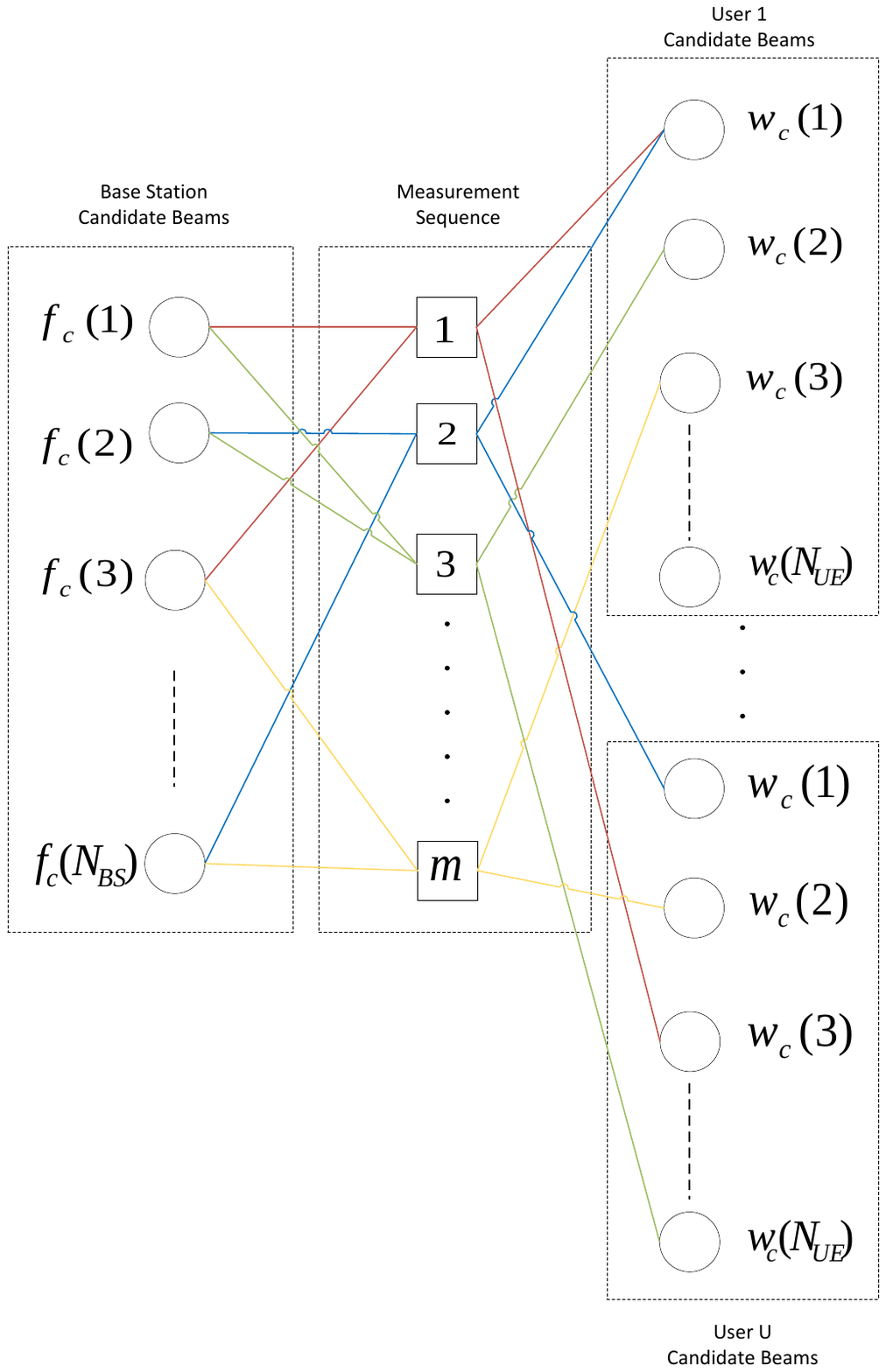}
\caption{Example graph-based model of how each channel measurement is comprised of random beam selections at the BS and each user. Solid black circles represent candidate beams at the BS and UEs, while the squares represent the increasing sequence of channel measurements.}
\label{MU_graph}
\end{figure}

In the proposed SWIFT framework, we transmit and receive with random combinations of these candidate beamforming vectors in order to estimate the channels of multiple UEs at the same time. \edit{By adopting a random sequence of transmit and receive directions, each user can continue to measure the channel for an arbitrary number of unique measurement observations}. We illustrate a graph-based model of how beams are selected in each measurement time slot in Fig. \ref{MU_graph}. For simplicity, the graph shows an example for the case that the BS selects two transmit candidate beams in each measurement time slot and each user selects just one receive candidate beam.

\subsection{Probabilistic Beam Selection for Channel Measurements}
We now can carry out channel measurements by adopting a sequence of randomly selected candidate beamforming vectors at both the BS and UEs. Specifically, in the $m$th measurement time slot, we propose to form $\bm{F}_m$ by randomly selecting $R_{\!B\!S}$ transmit candidate beamforming vectors from\footnote{Alternatively, a random number of beams may be employed in each measurement time slot, similar to concepts of the weight set and degree distribution in analog fountain codes \cite{shirvanimoghaddam2013near}. Here, to introduce the core idea of SWIFT, we utilize all RF chains in each channel measurement.} $\bm{F}_c$. Similarly, to form $\bm{W}_m^{(u)}$ at the $u$th user, we randomly select $R_{\!U\!E}$ receive candidate beamforming vectors from $\bm{W}_c$. Following (\ref{y}), we can then express the $u$th user's received signal in the $m$th measurement time slot as a $R_{\!U\!E}\times 1$ vector given by
\begin{align} \label{y_m}
\bm{y}_m^{(u)} &=  \sqrt{\frac{P}{R_{\!B\!S}}} (\bm{W}^{(u)}_m)^H  \bm{H}^{(u)}\bm{F}_m \bm{s}_m  + \bm{n}_m^{(u)}.
\end{align}
We first consider the simple case where equal probabilities of various candidate beams are used. Then the probability that the $n$th candidate vector $\bm{f}_c(n)$ is included in $\bm{F}_m$ at the BS becomes
\begin{align} \label{P_f_n}
\text{Pr}(\bm{f}_c(n)\in \bm{F}_m ) &=  \frac{R_{\!B\!S}}{N_{\!B\!S}}, \forall n=1,...,N_{\!B\!S}
\end{align}
and the probability that the $n$th candidate vector $\bm{w}_c(n)$ is included in $\bm{W}_m^{(u)}$ at the UE becomes
\begin{align} \label{P_w_n}
\text{Pr}(\bm{w}_c(n)\in \bm{W}_m^{(u)} )  &= \frac{R_{\!U\!E}}{N_{\!U\!E}}, \forall n=1,...,N_{\!U\!E}.
\end{align}
In all cases, we assume that the BS uses a pseudo-random number generator for the random beam selection process in each measurement and that this process can be predicted by each user, i.e., each UE knows which random beam selection the BS has made. In practice, this may require the BS to broadcast its pseudo-random seed before the first channel measurement of each new estimation process. Due to the low data requirement of this broadcast, the seed could be transmitted through a feedback channel while incurring little overhead.  

In this section, we consider the equal probability beam selection as described in (\ref{P_f_n}) and (\ref{P_w_n}) and modify the probabilities later in Section \ref{sec:nubp} to further improve the channel estimation performance. Fig. \ref{MU_graph} illustrates an example of a random beam selection process. As can be observed in the first measurement, the BS has selected both the first candidate beamforming vector $\bm{f}_c(1)$ and the third candidate beamforming vector $\bm{f}_c(3)$. In the same measurement, user 1 has selected  the first UE candidate beamforming vector $\bm{w}_c(1)$, while user U has selected $\bm{w}_c(3)$. We conclude this sub-section by expressing the sequence of all measurements up to the $m$th one collected at the $u$th user by a $ m R_{\!U\!E}\times 1$ vector given by

\begin{align} \label{y_all}
\bm{y}^{(u,m)} &=
\left[\begin{array}{c}
\bm{y}_1^{(u)} \\
\vdots   \\
\bm{y}_m^{(u)}
\end{array}\right]\\ &= \sqrt{\frac{P}{R_{\!B\!S}}}
\left[\begin{array}{c}
 (\bm{W}_1^{(u)})^H  \bm{H}^{(u)}\bm{F}_1 \bm{s}_1   \\
\vdots   \\
 (\bm{W}_m^{(u)})^H  \bm{H}^{(u)}\bm{F}_m \bm{s}_m
\end{array}\right] +
\left[\begin{array}{c}
\bm{n}_1^{(u)} \\
\vdots   \\
 \bm{n}_m^{(u)}
\end{array}\right].
\end{align}

\subsection{Sparse Estimation Problem Formulation}
In order to recover the virtual channel information using compressed sensing techniques, we require a standard-form expression \cite{vila2011expectation}, $\bm{y}^{(u,m)} = A_g\bm{A}^{(u,m)} \bm{v}^{(u)} + \bm{n}^{(u,m)}$, where $\bm{A}^{(u,m)}$ is an  $m R_{\!U\!E}\times N_{\!B\!S}N_{\!U\!E}$ sensing matrix, $A_g$ is a scalar constant, and $\bm{v}^{(u)}=\text{vec}(\bm{H}_v^{(u)})$ is the $N_{\!B\!S}N_{\!U\!E} \times 1$ vectorized virtual channel matrix to be detected.

To achieve a standard-form expression, we first rearrange (\ref{H_v}) by multiplying it by the left-hand pseudo inverse of $\bm{W}_c^H$ and right-hand pseudo inverse of $\bm{F}_c$ respectively. We then have
\begin{align} \label{H_H_v}
 \sqrt{N_{\!B\!S}N_{\!U\!E}}& \bm{W}_c(\bm{W}_c^H  \bm{W}_c)^{-1}\bm{H}_v^{(u)}  (\bm{F}_c^H \bm{F}_c)^{-1} \bm{F}_c^H  =  \\ & \bm{W}_c(\bm{W}_c^H\bm{W}_c)^{-1}  (\bm{W}_c)^H  \bm{H}^{(u)}\bm{F}_c (\bm{F}_c^H \bm{F}_c)^{-1} \bm{F}_c^H \nonumber
\end{align}
which, after algebraic manipulation, becomes
\begin{align}
      \bm{H}^{(u)} &=\sqrt{N_{\!B\!S}N_{\!U\!E}} \bm{W}_c\bm{H}_v^{(u)}   \bm{F}_c^H \label{H_H_v_3}
\end{align}

\noindent
where the simplification follows by the fact that $\bm{W}_c$ and $\bm{F}_c$ are matrices with orthogonal columns leading to $\bm{W}_c^H\bm{W}_c=\bm{I}_{N_{\!U\!E}}$ and $\bm{F}_c^H \bm{F}_c=\bm{I}_{N_{\!B\!S}}$. We can then substitute (\ref{H_H_v_3}) into (\ref{y_m}) to give
\begin{align} \label{y_m_H_v}
\bm{y}_m^{(u)} &=  \sqrt{\frac{PN_{\!B\!S}N_{\!U\!E}}{R_{\!B\!S}}} (\bm{W}_m^{(u)})^H  \bm{W}_c\bm{H}_v^{(u)}   \bm{F}_c^H \bm{F}_m \bm{s}_m  + \bm{n}_m^{(u)}.
\end{align}
By noticing that $\bm{y}_m^{(u)}$ is already a vector, we can then apply the property $\text{vec}(\bm{A}\bm{B}\bm{C}) = (\bm{C}^T \otimes\bm{A})  \text{vec}(\bm{B})$ to rewrite (\ref{y_m_H_v}) as
\begin{align} \label{y_m_H_v_vec}
\bm{y}_m^{(u)} \!  &= \! 
 A_g\big(   (\bm{F}_c^H \bm{F}_m \bm{s}_m)^T \!\! \otimes \!    (\bm{W}_m^{(u)})^H  \bm{W}_c  \big)   \text{vec}(\bm{H}_v^{(u)})     +   \bm{n}_m ^{(u)} \\
          \!  &= \!  A_g   \bm{A}_m^{(u)}   \text{vec}(\bm{H}_v)     +   \bm{n}_m^{(u)}  \label{y_m_H_v_vec_sub}
\end{align}
where $A_g = \sqrt{P N_{\!B\!S}N_{\!U\!E}/R_{\!B\!S}}$ and $\bm{A}_m^{(u)}=( \bm{s}_m^T \bm{F}_m^T \bm{F}_c^*)  \otimes   (  (\bm{W}_m^{(u)})^H  \bm{W}_c  ) $ is the $R_{\!U\!E}\times N_{\!B\!S}N_{\!U\!E}$ sensing matrix for the $m$th measurement. Finally, by substituting (\ref{y_m_H_v_vec_sub}) into (\ref{y_all}), we get
\begin{align} \label{y_all_CS_1}
\bm{y}^{(u,m)} &=  A_g
\left[\begin{array}{c}
\bm{A}_1^{(u)} \\
\vdots   \\
\bm{A}_m^{(u)}
\end{array}\right] \text{vec}(\bm{H}_v^{(u)}) +
\left[\begin{array}{c}
\bm{n}_1^{(u)} \\
\vdots   \\
 \bm{n}_m^{(u)}
\end{array}\right] \\
&=  A_g  \bm{A}^{(u,m)} \bm{v}^{(u)} + \bm{n}^{(u,m)}. \label{y_all_CS}
\end{align}

To complete the problem formulation, we now describe the statistics of each of the unknown terms in (\ref{y_all_CS}). In particular, we first focus on the virtual channel vector $\bm{v}^{(u)}$.

\edit{Although our channel model considers AOD/AOA that are distributed on the continuous angular range $[0,2\pi)$, for channel recovery purposes, the proposed estimation strategy approximates this model with a discrete set of angles that correspond to the directions of each candidate beam in (\ref{f_c})-(\ref{w_c}). As the set of candidate beamforming vectors together forms an orthogonal basis for the MIMO channel matrix, by estimating the virtual channel it is sufficient to reconstruct the channel itself.}  Physically, this quantization is the case where the AOD/AOA are perfectly aligned with each pair of the candidate beams. In this case, recalling $\alpha_l^{(u)}\sim\mathcal{C}\mathcal{N}(0,\sigma_R^{(u)})$, the channel sparsity can be characterized by a Bernoulli-Gaussian distribution, in which the $i$th entry of the vectorized virtual channel matrix $\bm{v}^{(u)}$ follows \cite{Jianhua_heath}
\begin{align} \label{v_i_prob}
{v_i^{(u)}} &\sim \begin{cases}
0 ,& \text{with probability  } 1- \rho^{(u)} \\
\mathcal{C}\mathcal{N}(0,\sigma_R^{(u)}) & \text{with probability  } \rho^{(u)}	
\end{cases}
\end{align}

\noindent
for all $i=1,\cdots,N_{\!B\!S}N_{\!U\!E}$ and $\rho^{(u)}= \text{E}[L^{(u)}]/(N_{\!B\!S}N_{\!U\!E})$ characterizes the degree of the channel sparsity of the $u$-th user.

We now turn our attention to the noise term $\bm{n}^{(u,m)}$ in (\ref{y_all_CS}). Recall from (\ref{y}) that the noise values, after being received with the set of beamforming vectors, follow distribution $ \mathcal{C}\mathcal{N}(0, N_0 (\bm{W}^{(u)})^H  \bm{W}^{(u)})$. As the adopted UE candidate beams are all mutually orthogonal to each other, the distribution of the $i$th element of $\bm{n}^{(u,m)}$ can then be simplified to
\begin{align} \label{n_i_prob}
{n}^{(u,m)}_i \sim   \mathcal{C}\mathcal{N}(0,N_0)
\end{align}
for all $i=1,\cdots,mR_{\!U\!E}$.

\begin{figure}[!t]
\centering
\includegraphics[width=3.5in,trim={0.9cm 8.9cm 3.0cm 1.0cm},clip]{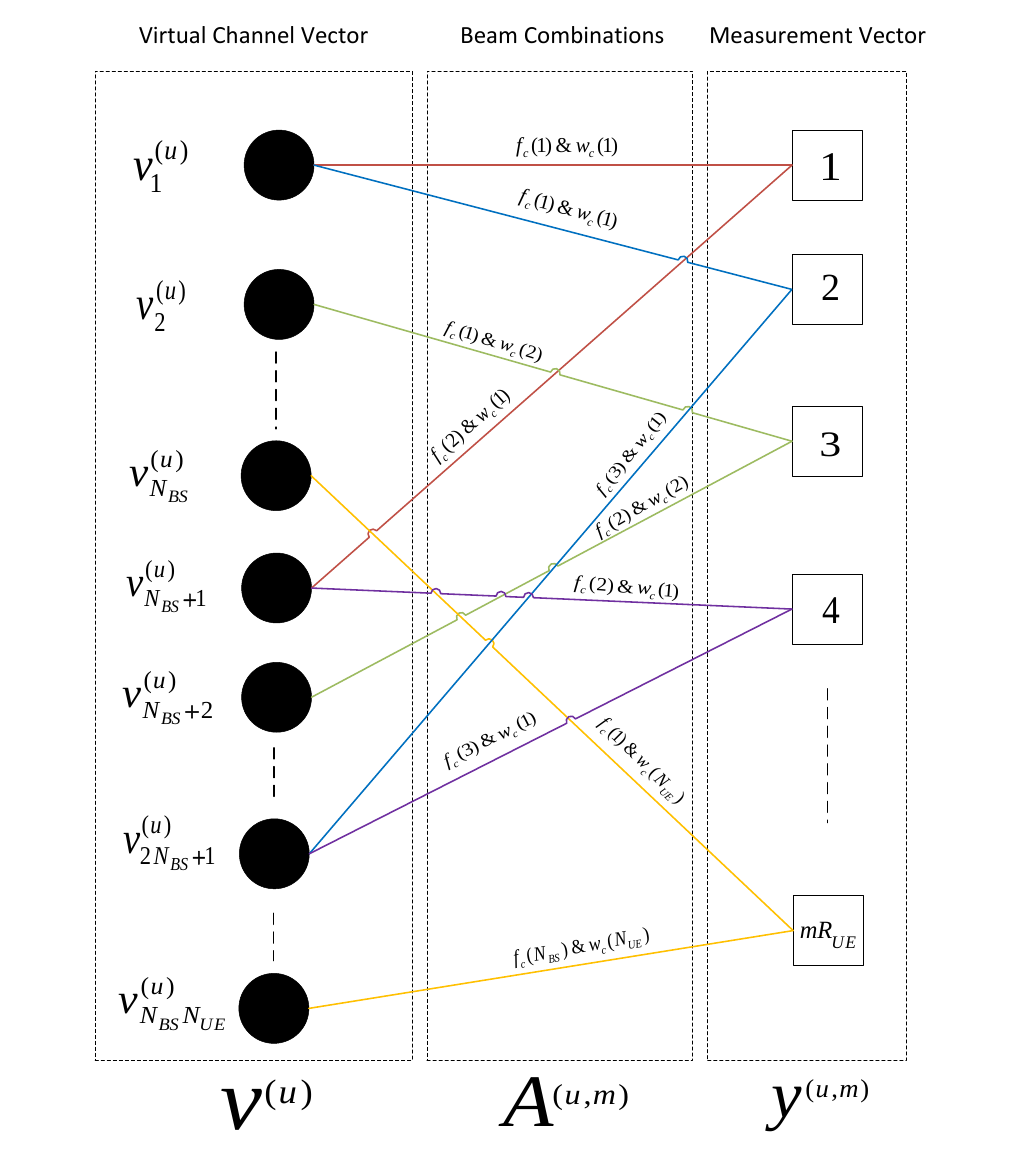}
\caption{Example graph-based model after formulating the compressed sensing problem.}
\label{CS_gprah}
\end{figure}

Since the channel estimation problem has now been formulated as a compressed sensing problem, the beam-selection graph in Fig. \ref{MU_graph} can be transformed to a bipartite graph, as shown in Fig. \ref{CS_gprah}. The variable nodes and check nodes shown on the left side and right side of Fig. \ref{CS_gprah} represent the virtual channel gains and measurement vectors, respectively. The links between the nodes depict the random beam selection characterized by the sensing matrix $\bm{A}^{(u,m)}$. Fig. \ref{CS_gprah} therefore elaborates on the physical relationship between the measurements and the channel, which can be expressed as $\bm{y}^{(u,m)} = A_g \bm{A}^{(u,m)} \bm{v}^{(u)}$. For example, it can be seen in Fig. \ref{CS_gprah} that the first measurement is generated by adopting the combinations $\bm{f}_{c}(1)$ with $\bm{w}_{c}(1)$ and $\bm{f}_{c}(2)$ with $\bm{w}_{c}(1)$. More generally, candidate beam pair $\bm{f}_{c}(i)$ and $\bm{w}_{c}(j)$ can be seen to link to the virtual channel vector index ${v_{(i-1)N_{\!U\!E}+j}^{(u)}}$. Each new measurement will create an additional check node on the right side, linking more variable nodes together. Similar bipartite graphs can be seen in the design of analog fountain codes \cite{shirvanimoghaddam2013near}. Such a graph representation of channel estimation process enables us to apply the mature code-on-graph theory to tackle the challenging channel estimation problem and apply powerful message passing decoding algorithms for channel information recovery in the following sub-section.

\subsection{UE Virtual Channel Information Recovery Using GAMP}
\label{sssec:sec_gamp}
We now need a method to efficiently estimate the virtual channel information in $\bm{v}^{(u)}$, based on the measurements $\bm{y}^{(u,m)}$ at each user. A maximum likelihood solution to our estimation problem can take the form
\begin{align} \label{ML_prob}
\hat{\bm{v}}^{(u,m)} &= \underset{\bm{v}}{\operatorname{argmax}} [ p(\bm{y}^{(u,m)}|\bm{v}) ].
\end{align}
where $\hat{\bm{v}}^{(u,m)}$ is the $u$th user's estimate of the virtual channel vector, based upon all measurements obtained after $m$ time slots. Unfortunately, the general maximum likelihood estimator does not consider the sparsity of $\bm{v}^{(u)}$. It has been shown in \cite{tibshirani1996regression} that the Lasso outperforms maximum likelihood in sparse estimation by leveraging the inherent sparsity. From a probabilistic view, the Lasso estimator is equivalent to the maximum likelihood one under the assumption that the entries of the estimated vector follow a Laplace distribution \cite{huang2016approximate}. When it comes to our channel estimation problem, the Lasso estimator will solve the following problem
%
\begin{align} \label{Lasso}
\hat{\bm{v}}^{(u,m)} &= \underset{\bm{v}}{\operatorname{argmin}}\Big[ || \bm{y}^{(u)} -  A_g \bm{A}^{(u,m)} \bm{v}||_2^2 + \gamma || \bm{v} ||_1 \Big]
\end{align}
where $|| \bm{y}^{(u)} - A_g \bm{A}^{(u,m)} \bm{v}||_2^2$ is the data-promoting term to ensure the estimate fits the observations, $|| \bm{v} ||_1$ is the sparsity-promoting term, which essentially reduces the number of non-zero values in the solution, and $\gamma$ balances the tradeoff between these two terms. However, finding optimal solutions to (\ref{Lasso}) generally becomes computationally expensive when the dimension of the estimated vector is sufficiently large. Motivated by this, generalized approximate message passing approximation (GAMP) solutions have been developed in \cite{rangan2011generalized} to approximate (\ref{Lasso}).


The general idea of GAMP is to find the approximate solution to (\ref{Lasso}) by taking into consideration the channel statistics (e.g., sparsity). In \cite{rangan2011generalized}, GAMP is proposed for arbitrary channel statistics and characterized by two functions $g_{out}(*)$ and $g_{in}(*)$. These two functions essentially describe the statistics of the estimation input and output vectors. In our case, $g_{in}(*)$ describes estimation input which refers to our Bernoulli-Gaussian virtual channel vector with statistics described in (\ref{v_i_prob}). Similarly, $g_{out}(*)$ characterizes the estimation output which refers to the complex AWGN channel output vector with statistics described in (\ref{n_i_prob}). With these statistics, we can adopt the Bernoulli-Gaussian GAMP estimator described in \cite{vila2011expectation} to estimate the virtual channel vector $\bm{v}^{(u,m)}$. \edit{We assume that the statistics of the propagation paths, such as the average number and path gains, are known to the receiver. In practice, this could be obtained from previous rounds of channel estimation. In particular, as the average channel sparsity $\rho^{(u)}$ is only expected to vary over a large time-scale, (e.g., urban, rural) and could therefore be gradually updated over many channel realizations.}

Due to space limitation, we omit the detailed discussion of the GAMP algorithm and instead refer interested readers to \cite{rangan2011generalized,vila2011expectation}. For the completeness and reproducibility, we formally describe the Bernoulli-Gaussian GAMP estimator used in this paper in Algorithm 1. The GAMP estimator can be seen to iteratively update its estimate until it converges on a final output denoted by $\hat{\bm{v}}^{(u,m)}$.  We use this estimate in the following sub-sections to give a complete description of SWIFT.

%

\subsection{UE Stopping Criterion}

As the proposed BS beam patterns do not adapt to any particular user, our method is able to simultaneously estimate all downlink channels for multiple users. We propose that the BS continues to transmit pilot signals with randomly selected beamfoming vectors, until each user's channel estimation has accurately converged. We follow the methods used in sequential compressed sensing \cite{malioutov2010sequential} and consider the estimate complete when the current estimate has not changed significantly from the previous one. To implement this approach in our framework, recalling (\ref{v_i_prob}), we binarize the estimated virtual channel vector as
\begin{align} \label{v_gamma}
\bar{{v}}_i^{(u,m)} &= \begin{cases}
    	  0 ,& \text{if } |\hat{{v}}_i^{(u,m)}|< \Gamma \sigma_R^{(u)} \\
    	  1 ,& \text{otherwise}	
		 \end{cases}
\end{align}
where $\Gamma<<1$ determines the threshold of path coefficients that can be considered to be negligible or in a deep fade\footnote{In practice, $\Gamma$ could be set according to the minimum fading coefficient that the transceiver can use for acceptable communication, and would depend on the required rate of the system, transmit power etc.}. We then consider that the channel estimate has converged if the new binarized virtual channel vector is equal to the previous one. That is, the channel estimation of the $u$th user is deemed as complete if $\bar{\bm{v}}^{(u,m)}=\bar{\bm{v}}^{(u,m-T_u)}$, where $T_u$ determines how many measurements are carried out between GAMP estimation updates. We define the time in terms of symbols required for the $u$th user to reach this stopping criterion as $T_E^{(u)}$. To prevent an infinite sequence of measurements when the channel is in a deep fade or completely blocked, we introduce a maximum allowed number of measurements, denoted by $T_{max}$. 

\begin{figure}[!t]
\removelatexerror
\begin{algorithm}[H]
\label{alg2}
\caption{Bernoulli-Gaussian Generalized Approximate Message Passing (GAMP) Algorithm from \cite{vila2011expectation}.}
{\fontsize{9}{9}\selectfont
$\bm{Input:}$ $ \bm{y}^{(u,m)} \rightarrow \bm{y}$, $  \bm{A}^{(u,m)} \rightarrow \bm{A} $, $A_g$,  $\rho^{(u)}$, $\sigma_R^{(u)}$ and $N_0$. \vspace{0.5pt}  \\
$\bm{Initialization:}$  $\hat{\bm{v}}^{(1)}=\bm{0}$, $\text{Var}[\hat{\bm{v}}^{(1)}]=\bm{1}$ and $\hat{\bm{s}}^{(0)}=\bm{0}$ \vspace{0.5pt} \\
$\bm{Define:}$ 	${a}_{i} \text{ is the $i$th entry of vector $\bm{a}$}.$\\	
$A_{i,j} \text{ is the entry on the $i$th row and $j$th column of the matrix $\bm{A}$}.$ 
\begin{align}
\bm{Define}&\bm{\text{ }Characteristic \text{ }Functions:}& \vspace{-10.1pt}&	\nonumber \\
&{g}_{out}(y,\hat{p},\text{Var}[\hat{p}]) = \frac{y-\hat{p}}{\text{Var}[\hat{p}]+ N_0} \nonumber  \hspace{16cm} \\
-&{g}_{out}'(y,\hat{p},\text{Var}[\hat{p}]) = \frac{1}{\text{Var}[\hat{p}]+ N_0}, \nonumber  \\
&{g}_{in}(\hat{r},\text{Var}[\hat{r}]) = \pi(\hat{r},\text{Var}[\hat{r}]) \gamma(\hat{r},\text{Var}[\hat{r}]) \nonumber  \\
-\text{Var}[\hat{r}]&{g}_{in}'(\hat{r},\text{Var}[\hat{r}]) = \pi(\hat{r},\text{Var}[\hat{r}]) \big(\nu(\hat{r},\text{Var}[\hat{r}]) + |\gamma(\hat{r},\text{Var}[\hat{r}])|^2 \big) \nonumber \\
&\hspace{3cm}  -(\pi(\hat{r},\text{Var}[\hat{r}]))^2|\gamma(\hat{r},\text{Var}[\hat{r}])|^2, \nonumber  \\
\text{where}  &  \nonumber\\
&\pi(\hat{r},\text{Var}[\hat{r}]) \triangleq\frac{1}{1+ \frac{1-\rho^{(u)}}{\rho^{(u)}} \frac{ \mathcal{C}\mathcal{N}(\hat{r},0,\text{Var}[\hat{r}] ) }{\mathcal{C}\mathcal{N}(\hat{r},0,\text{Var}[\hat{r}]+ \sigma_R^{(u)})} }, \nonumber  \\
&\gamma(\hat{r},\text{Var}[\hat{r}]) \triangleq \frac{\hat{r}/\text{Var}[\hat{r}]}{1/\text{Var}[\hat{r}] + 1/\sigma_R^{(u)}}, \text{ and } \nonumber \\
&\nu(\hat{r},\text{Var}[\hat{r}]) \triangleq \frac{1}{1/\text{Var}[\hat{r}] + 1/\sigma_R^{(u)}}. \nonumber
\end{align}
// Begin Estimation  \vspace{0.5pt} \\
   \For( \emph{}){$k=1,2,...$}
   {    \vspace{2.5pt} // Output linear step. \vspace{2.5pt} \\
    $\hat{{z}}_i^{(k)}  =  \sum_j A_g {A}_{i,j} \hat{{v}}_j^{(k)}$  $\forall i$  \vspace{0.5pt} \\
    $\text{Var}[\hat{{z}}_i^{(k)}]   =  \sum_j |A_g {A}_{i,j}|^2 \text{Var}[{{v}}_j^{(k)}]$  $\forall i$     \vspace{5.0pt}\\    	
    // Output non-linear step.\vspace{2.5pt} \\
    $ \hat{{s}}^{(k)}_i   = g_{out}( {y}_i, \hat{{z}}^{(k)}_i - \text{Var}[\hat{{z}}^{(k)}_i]\hat{{s}}^{(k-1)}_i      ,\text{Var}[\hat{{z}}^{(k)}_i]) $  $\forall i$ \vspace{0.5pt}\\
    $ \text{Var}[\hat{{s}}^{(k)}_i]   = -g'_{out}( {y}_i, \hat{{z}_i}^{(k)} - \text{Var}[\hat{{z}}^{(k)}_i]\hat{{s}}^{(k-1)}_i , \text{Var}[\hat{{z}}^{(k)}_i] )$  $\forall i$ \vspace{0.0pt}\\
   // Input linear step. \vspace{2.5pt} \\
   $ \text{Var}[\hat{{r}}^{(k)}_j]  = 1/(  \sum_i |A_g {A}_{i,j}|^2 \text{Var}[\hat{{s}}^{(k)}_i] ) $  $\forall j$	\vspace{0.5pt}	\\
   $ \hat{{r}}^{(k)}_j  = \hat{{v}}^{(k)}_j + \text{Var}[\hat{{r}}^{(k)}_j] \sum_i A_g {A}_{i,j}^* \hat{{s}}^{(k)}_i  $  $\forall j$	\vspace{5.0pt}	\\
   // Input non-linear step. \vspace{2.5pt}  \\
   $ \hat{{v}}^{(k+1)}_j =   g_{in}( \hat{{r}}^{(k)}_j, \text{Var}[\hat{{r}}^{(k)}_j]  )  $  $\forall j$		\vspace{0.5pt} \\
   $ \text{Var}[\hat{{v}}^{(k+1)}_j] =   -\text{Var}[\hat{{r}}^{(k)}_j] g'_{in}( \hat{{r}}^{(k)}_j, \text{Var}[\hat{{r}}^{(k)}_j] )   $  $\forall j$		\vspace{5.0pt} \\
   // Check for convergence. \vspace{2.5pt} \\
   \If( \emph{}){$\hat{\bm{v}}^{(k+1)}=\hat{\bm{v}}^{(k)}$}
   		{
   		break
		}
   }
$\bm{Output:}$ $ \hat{\bm{v}}^{(k+1)} \rightarrow \hat{\bm{v}}^{(u,m)}$.
}
\end{algorithm}
\end{figure}

\subsection{UE Beam Selection for Data Communication}
After meeting the channel estimation stopping criterion, the user stops its estimation process and feeds back the indices of beamforming vectors to be adopted by the BS for the ensuing data communication. To determine these beamforming indices, after each user converts the estimated channel vector $\hat{\bm{v}}^{(u,T_E^{(u)})}$ back into its matrix form (i.e., $\hat{\bm{{H}}}_v^{ { { (u,T_E^{(u)}) } } }$), the user then determines the candidate beams (for both the BS and UE) that maximize the achievable rate. Recalling the transceiver relationship given in (\ref{x})-(\ref{y}), this involves finding a BS beamforming matrix, $\bm{F}_d$, and user beamforming matrix, $\bm{W}_d$, that maximizes the achievable rate of the $u$th user given by \cite{rheath}
\begin{align} \label{Rate}
R_{opt}^{(u)}= \text{log}_2|\bm{I} + \frac{P}{N_0} \bm{W}_d^H \hat{\bm{H}}^{(u,T_E^{(u)})} \bm{F}_d \bm{F}_d^H \hat{\bm{H}}^H \bm{W}_d|.
\end{align}
Recalling from (\ref{H_H_v}) that $\bm{H}^{(u,T_E^{(u)})}= \sqrt{N_{\!B\!S}N_{\!U\!E}} \bm{W}_c\bm{H}_v^{(u,m)}   \bm{F}_c^H$, we then have

\begin{align} \label{Rate_opt}
\{ \bm{F}_{opt}^{(u)},\bm{W}_{opt}^{(u)} \} &=  \underset{\bm{F}_d,\bm{W}_d}{\operatorname{argmax}}   \text{ log}_2|\bm{I} +  \\  A_g^2 \bm{W}_d^H \bm{W}_c&\hat{\bm{H}}_v^{(u,T_E^{(u)})}\bm{F}_c^H \bm{F}_d \bm{F}_d^H \bm{F}_c  (\hat{\bm{H}}_v^{(u,T_E^{(u)})})^H \bm{W}_c^H \bm{W}_d|. \nonumber
\end{align}
%
As the columns of the communication beamforming matrices can only consist of candidate beamforming vectors, $\bm{F}_d$ and $\bm{W}_d$ are constrained to finite set of vectors. \edit{Furthermore, due to the mutual orthogonality among the candidate beams selected in ${\bm{F}_d}$ and ${\bm{W}_d}$, the matrix multiplication terms $\bm{F}_c^H \bm{F}_d$ and $\bm{W}_c^H \bm{W}_d$, along with their conjugate transposes, can be expressed as sparse matrices, with each non-zero values on the diagonal entries corresponding to a selected candidate beam index. As such, the optimal beam selection can be reduced to simply finding the row and column indices of the virtual channel estimate that maximizes the logarithm term in (\ref{Rate_opt}).}

 (\ref{Rate_opt}) can be reduced to finding the indexes of the largest magnitude values in $\hat{\bm{H}}_v^{(u,T_E^{(u)})}$.


Due to the limited feedback bandwidth in the multi-user scenario, we consider that each user is only able to feedback the BS-side beamforming directions determined by (\ref{Rate_opt}), and not the path fading coefficient. However, it is worth pointing out that the path fading coefficient is still used for coherent detection at the UE side. As such, we consider that the BS allocates equal power to all identified paths. This reduces the number of feedback bits to only $\lceil\text{log}_2(N_{BS})\rceil$ per estimated path. To characterize the performance of the proposed SWIFT algorithm, we follow \cite{alkhateeby2015compressed} and define the effective rate of the $u$th user, given the time ratio consumed for the channel estimation, by
\begin{align} \label{effective_rate}
 R_{E}^{(u)} =  R_{opt}^{(u)} \big(1 - \frac{T_E^{(u)}}{T_c}\big),
\end{align}
recalling that $T_c$ is the coherence time of each channel realization.

\subsection{BS Stopping Criterion and User-scheduling}
\label{sec:BS_stop}
We consider two scenarios for the BS stopping criterion of channel estimation, i.e., when the BS is to stop broadcasting pilot symbols and commence data communication. The first one is the ideal case where the BS can perform data communication with users in adjacent sub-channels. In this case, we propose that once a user believes that it has completed its estimation and feeds back the beamforming directions, the BS will use the feedback information and start to communicate with this user using an adjacent sub-channel straight away. The BS can continue to broadcast pilot signals on the previous sub-channel for other users that have not finished their channel estimation. Similar out-of-band estimation approaches have also been proposed in \cite{nitsche2015steering}. As the relative change in frequency for using an adjacent sub-carrier is quite low in the mmWave band, it is reasonable to assume that the AOD/AOA directions remain unchanged in the adjacent sub-carrier, although we acknowledge that in practice a few initial pilots may be required in the new sub-channel to refine the estimate of the fading coefficient at the user side. Extension to time and spatial domain multiplexing may also be possible as the BS coordinates the usage of all beamforming directions among multiple users.


In the second case, we consider the BS and UE channel estimation and data communication to occur in the same frequency band and communicate in different time intervals. In this case, we propose that the BS can perform user scheduling by leveraging SWIFT's capability of inferring the channel quality sequences of multiple users based on the sequences that the users finish their channel estimation. Specifically, those users that complete (feedback) their channel estimation earlier normally have better channel conditions than those finish the channel estimation later, and thus are more suitable to communicate with BS in the current channel realization. In this sense, once the BS has collected $N_{s}\leq U$ user's channel feedbacks, it can stop broadcasting pilot symbols and begin communication with selected users. \edit{As we show later in our numerical results (particularly, \ref{P_CDF}. 8), users who complete their channel estimation early are expected to have better channels than those who finish later. Intuitively speaking, this follows from the stability of the estimation process in different SNR ranges. Based on this observation, it can be deduced that UE with better channel conditions will, on average, compete their channel estimation before those with worse channel conditions.}

%
\subsection{A Summary of SWIFT}

\begin{figure}[!t]
\centering
\includegraphics[width=2.9in,trim={0.6cm 2.4cm 2.0cm 1.05cm},clip]{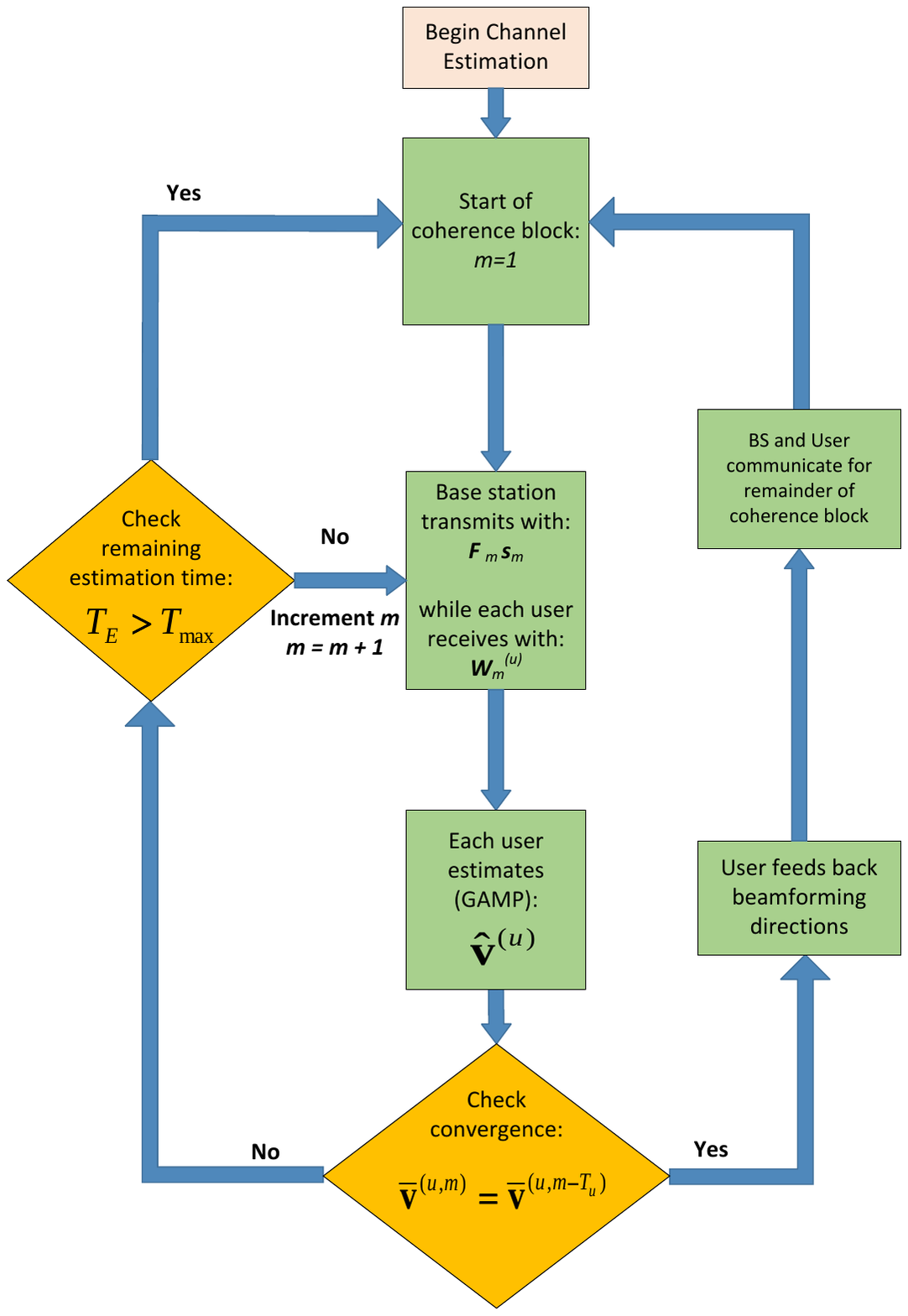}
\caption{Channel estimation flow diagram for each user in the proposed SWIFT framework.}
\label{flow_diagram}
\end{figure}

We are now ready to summarize the proposed SWIFT framework. To this end, we provide a flow diagram of the complete SWIFT algorithm at user side in Fig. \ref{flow_diagram}. We also elaborate each step in SWIFT as follows:

\begin{enumerate}
\setlength{\itemindent}{0.145in}
\item[Step 1:] In each measurement time slot, the BS randomly selects $R_{BS}$ candidate beamforming vectors to transmit the pilot signals. At the same time, each user randomly selects $R_{\!U\!E}$ candidate beamforming vectors to receive the pilot signals.
\item[Step 2:] Each user implements the GAMP algorithm to estimate its channel information based on all the collected measurements until the current time slot.
\item[Step 3:] If the estimated channel has converged to the predefined accuracy or if the maximum estimation time $T_{max}$ has been reached, the channel estimation is considered to be complete and this user can proceed to Step 4. Otherwise go back to Step 1.
\item[Step 4:] The user determines the optimal beamforming vectors to be used for data communication and feeds back the beamforming indices for the BS to perform data transmission in the remaining $T_c-T_E$ time slots.
\end{enumerate}
At the beginning of each transmission block, the process returns to Step 1 and repeats. We end this section by highlighting several key benefits of the proposed SWIFT scheme as follows:
\begin{itemize}
\item{Due to the stochastic nature of when each user completes its channel estimation, user feedback events are distributed randomly throughout the whole estimation procedure, resulting in less pressure on the bandwidth of feedback channels. }
\item{\edit{As our algorithm is inherently designed for various channel estimations with different estimation times, the extension to include a range of different number of antennas and RF chains at the UEs is straightforward. However, in this case, the proposed user-scheduling scheme would also need to consider this to provide fairness toward users with more complex MIMO channels.}}
\item{As the time occurrence of user feedbacks gives an insight into channel quality, without any additional feedback other than directions of paths. This implicit channel quality information could be leveraged to achieve certain QoS requirements.}
\item{The probabilistic feature of the beam selection naturally allows any prior/partial channel knowledge to be applied to improve channel estimation performance, e.g., allocating a higher probability of beam selection to beams nearer to the previously identified AOD/AOA. }
\end{itemize}

\noindent

\section{Non-uniform Beam Probabilities}
\label{sec:nubp}
Inspired by the concept of unequal error protection in fountain codes, in this section we propose two modifications to the initially defined uniform beam probabilities in (\ref{P_f_n})-(\ref{P_w_n}). To proceed, we define the vector $\bm{\delta}^{(m)}=[\delta^{(m)}_1,\cdots,\delta^{(m)}_n,\cdots, \delta^{(m)}_{N_{\!B\!S}}]$ to describe the probability of each candidate beam being selected for use by the first RF chain in $m$th measurement\footnote{It is worth noting that the probability of each beam being selected for use with subsequent RF chains is impacted by the beams that have been selected  previously and therefore cannot be selected again. This leads to a ``weighted random selection without replacement'' process.}. Similarly, we denote the beam selection probability vector at the $u$th user for the $m$th measurement as $\bm{\epsilon}^{(u,m)}=[\epsilon^{(u,m)}_1,\cdots,\epsilon^{(u,m)}_n,\cdots, \epsilon^{(u,m)}_{N_{\!U\!E}}]$. The first modification is used to avoid the case that a given beam combination is not selected at least once before $T_{max}$. This is done by introducing a forcing approach that decreases the average number of measurement time slots required to span all beam combinations at least once. In the second modification, we propose a user-side partially estimated probability adaptation (PEPA) scheme to adjust the beam probabilities based upon the estimated channel available in the previous time slot.

\subsection{Forcing Probability Adaptation (FPA)}
\label{ssec:sec_forcing}
In this subsection, we address the non-zero probability that a given transmit and receive beam combination is not spanned at least once before the maximum estimation time has been reached. To this end, after each measurement time slot, we propose to set the BS beam selection probability vector for the next measurement to be inversely proportional to the total number of times that a given beam has already been selected. We denote ${N}_f^{(m)}(n)$ as the number of times the $n$th candidate beamforming vector $\bm{f}_c(n)$ has been used at the BS after the $m$th measurement. We can then express the beam selection probability vector for the $(m+1)$th measurement as
\begin{align} \label{f_forcing}
\bm{\delta}^{(m+1)}=  \left[\frac{c_\delta}{{N}_f^{(m)}(1)+ \eta},\frac{c_\delta}{{N}_f^{(m)}(2)+ \eta},\cdots,\frac{c_\delta}{{N}_f^{(m)}(N_{\!B\!S})+ \eta}\right].
\end{align}
where $c_\delta=(\sum_n( {N}_f^{(m)}(n) + \eta )^{-1})^{-1}$ is a scalar constant that ensures that the sum of the entries in $\bm{\delta}^{(m+1)}$ add to one and $\eta$ is a sufficiently small positive value that prevents the occurrence of a zero denominator. It is worth noting that, although (\ref{f_forcing}) affects which BS beams are selected at the BS, it does not depend on any information that is not known by each user. As such, each user can still predict the beam selection at the BS for each subsequent measurement.

On the user-side, this type of adaptation is not as straightforward as in (\ref{f_forcing}). This is because no user can affect the beam selection at the BS. To increase the chance that each candidate beam combination is spanned at least once, each user should ensure that there is at least one non-zero entry in each column of the sensing matrix, $\bm{A}^{(u,m)}$. As such, each user should take into consideration the BS beams to be used in the next time slot (i.e., $\bm{F}_{m+1}$) when modifying its beam selection probability. Specifically, we denote ${N}_w^{(m)}(n|\bm{f}_c)$ as the number of times that the $n$th candidate beam $\bm{w}_c(n)$ has been used at the UE in conjunction with candidate beam $\bm{f}_c$ being used at the BS, after the $m$th measurement time slot. We can then propose to update the beam selection probability vector at the $u$-th user for the $(m+1)$th measurement as
\begin{align} \label{w_forcing}
\bm{\epsilon}^{(u,m+1)}=\;\;\;\;\;\;\;\;\;\;\;\;\;\;\;\;\;\;\;\;\;\;\;\;& \nonumber \\    \Bigg[   \frac{c_\epsilon}{  \underset{\bm{f}_c\in \bm{F}_{m+1} }{\operatorname{min}}   {N}_w^{(m)}(1|\bm{f}_c)  + \eta }&, \cdots,\frac{c_\epsilon}{ \underset{ \bm{f}_c \in \bm{F}_{m+1} + \eta}{\operatorname{min}} {N}_w^{(m)}(N_{\!B\!S}|\bm{f}_c ) + \eta   }\Bigg].
\end{align}
where $c_\epsilon$ is a scalar constant that ensures that the sum of the entries in $\bm{\epsilon}^{(m+1)}$ add to one. Equation (\ref{w_forcing}) essentially sets the probability of each UE candidate beamforming vector in the next time slot according to the number of times it has been used together with the candidate beams that are about to be adopted by the BS. The ``min'' operation emphasizes the BS candidate beam that has been used with each UE candidate beam the least. By adopting the FPA approach as described by (\ref{f_forcing}) and (\ref{w_forcing}), the average number of measurement time slots required to span all beam combinations can be significantly reduced, compared to the default scheme with uniform beam probabilities. Note that similar forcing strategies are normally applied in fountain codes to avoid an error floor at high SNR \cite{hussain2011error}.

%
\subsection{Partially Estimated Probability Adaptation (PEPA)}
\label{ssec:sec_pepa}
In this subsection, we propose to exploit the estimated virtual channel matrix obtained from all previous measurements, to increase the power of the received signal in the subsequent measurements. To achieve this, we propose that once all beam possible combinations have been spanned at least once, each user modifies its beam selection probabilities based on its recently estimated channel information, which is referred to as partially estimated probability adaptation (PEPA) in this paper. This information can be used in such a way to maximize the received signal power and therefore maximize the amount of channel information carried by the signal. In particular, we note that after time slot $m$, the user knows the beamforming matrix to be used by the BS in the next measurement time slot (i.e., $\bm{F}_{m+1}$) and also has an estimate of the channel based on all previous measurements $\hat{\bm{H}}^{(u,m)}$. Based on these two important pieces of information, each user can then make an estimate of the signal to be received by each antenna in the next time slot. From (\ref{r}) we then can express the signal to be received for the $(m+1)$th measurement as
\begin{align} \label{r_est}
\hat{\bm{r}}_{m+1}^{(u)} &= \sqrt{\frac{P}{R_{\!B\!S}}}  \hat{\bm{H}}^{(u,m)}\bm{F}_{m+1} \bm{s}_{m+1}.
\end{align}
Using this prediction, each user can then estimate the expected received measurement given the $n$th candidate beamforming vector as $(\bm{w}_c(n))^H  \hat{\bm{r}}_{m+1}^{(u)}$. To maximize the expected signal power in the next time slot, we then propose to update the beam probabilities for each user in the next time slot proportional to the expected signal power for each candidate beamforming vector. Mathematically, we have

\begin{align} \label{w_opt}
\bm{\epsilon}^{(u,m+1)} = c_\epsilon  \Big[ &  (\bm{w}_c(1))^H  \hat{\bm{r}}_{m+1}^{(u)} (\hat{\bm{r}}_{m+1}^{(u)})^H \bm{w}_c(1), \cdots, \nonumber \\ &\;\;\;\;\; (\bm{w}_c(N_{ \!U\!E}))^H  \hat{\bm{r}}_{m+1}^{(u)} (\hat{\bm{r}}_{m+1}^{(u)})^H \bm{w}_c(N_{ \!U\!E})] \Big] \\
=   \; & c_\epsilon \text{diag}\Big(\bm{W}_c^H \hat{\bm{r}}_{m+1}^{(u)} (\hat{\bm{r}}_{m+1}^{(u)})^H  \bm{W}_c \Big).
\label{w_opt_1}
\end{align}
Substituting (\ref{r_est}) into (\ref{w_opt_1}) and recalling from (\ref{H_H_v_3}) that $ \bm{H}^{(u)}= \bm{W_c}\bm{H}_v^{(u)}  \bm{F_c}^H$, we then have
\begin{align} \label{w_opt_2}
\bm{\epsilon}^{(m+1)}  &=   c_\epsilon A_g \text{diag}\Big(\bm{W}_c^H   \hat{\bm{H}}^{(u,m)}\bm{F}_{m+1} \bm{s}_{m+1} \nonumber \\ &\;\;\;\;\;\;\;\;\;\;\;\;\;\;\;\;\;\;\;\;\;\;\;\; (  \hat{\bm{H}}^{(u,m)}\bm{F}_{m+1} \bm{s}_{m+1})^H  \bm{W}_c \Big) \\
 &=  c_\epsilon A_g \text{diag}\Big(\bm{W}_c^H  \bm{W}_c \hat{\bm{H}}_v^{(u,m)}  \bm{F}^H_c  \bm{F}_{m+1} \bm{s}_{m+1} \bm{s}_{m+1}^H \nonumber \\&\;\;\;\;\;\;\;\;\;\;\;\;\;\;\;\;\;\;\;\;\;\;\;\;   \bm{F}_{m+1}^H   ( \bm{W_c} \hat{\bm{H}}_v^{(u,m)} \bm{F}_c   )^H   \bm{W}_c \Big) \\
 &= c_\epsilon A_g \text{diag}\Big(\hat{\bm{H}}_v^{(u,m)} \bm{Q}_{m+1} (\hat{\bm{H}}_v^{(u,m)})^H   \Big)
\end{align}
where the matrix $\bm{Q}_{m+1} = \bm{F}^H_c  \bm{F}_{m+1} \bm{s}_{m+1} \bm{s}_{m+1}^H \bm{F}_{m+1}^H  \bm{F}_c^H $ is a sparse diagonal matrix with only $R_{BS}$ non-zero elements.

By using the PEPA approach, users are able to utilize partially estimated channel information to achieve a stronger estimate of the channel in the subsequent measurements. This can be considered analogous to the concept of unequal error protection through intermediate feedback in fountain codes \cite{abbas2016performance}. Unlike fountain codes, as our beam combinations are jointly determined by both link ends, we are able to implement this adaptive concept of ``unequal beam protection'' without incurring any additional feedback overhead. It is also worth noting that in the single user regime, the use of partial channel feedback during the estimation process may enable the BS to also adapt its beam probabilities to maximize the signal power to be delivered to the user. As we have mainly focused on a multi-user scenario in this paper, we do not consider this BS-side beam adaptation here. However, in the multi-user regime, unequal beam protection at the BS may resemble to Fountain Codes in multi-cast scenarios \cite{ahmad2010adaptive}, which is out of the scope of this paper and has been left as a future work.

\section{Convergence Analysis}
To verify whether the proposed algorithm can converge for any arbitrary number of antennas and RF chains, we can consider whether the corresponding measurement matrix, \(\bm{A}^{(u,m)}\) satisfies the reconstruction criterion for the GAMP estimation. Specifically, if the matrix satisfies the restricted isometric property (RIP) then it can be guaranteed that the channel can be reconstructed by L1-minimization such as GAMP \cite{foucart2013mathematical}. Following \cite{foucart2013mathematical} we define the \emph{restricted isometry} \emph{constant} \(\delta_{L}\) of matrix
\(\bm{A}^{\left( u,m \right)}\ \) as the smallest \(\delta_{L} \geq 0\) that satisfies
\begin{align}
\left( 1 - \delta_{L} \right)\left\| \bm{v}^{\bm{(}u\bm{,}m\bm{)}} \right\|_{2}^{2} &\leq \left\| \bm{A}^{\left( u,m \right)}\bm{v}^{\left( u\bm{,}m \right)} \right\|_{2}^{2}\ \nonumber \\ &\leq \left( 1 + \delta_{L} \right)\left\| \bm{v}^{\left( u\bm{,}m \right)} \right\|_{2}^{2}\ ,\forall \; \bm{v}^{\left( u\bm{,}m \right)}\
\end{align}
across all possible \(L\)-sparse UE channels, \(\bm{v}^{\bm{(u)}}\) satisfying
\(\left\| \bm{v}^{\left( u\bm{,}m \right)} \right\|_{0} = L\).
Informally, the matrix \(\bm{A}^{\left( u,m \right)}\ \)is then said
to possess the RIP if \(\delta_{L}\ \)is small for sufficiently large
\(L\). More rigorously, the RIP will be satisfied if all sub matrices
formed by \(L\) columns of \(\bm{A}^{\left( u,m \right)}\) are well
conditioned.

Unfortunately, the evaluation of RIP across all possibilities is a non-trivial problem \cite{bandeira2013certifying}. In addition to the stochastic process in \(\bm{A}^{\left( u,m \right)}\), the considered model also has random sparsity in the number paths with each path also following a Gaussian distribution. This makes it difficult to precisely evaluate the RIP across all possible random sensing matrices \(\bm{A}^{\left( u,m \right)}\), although it is widely accepted that
random matrices are a good choice \cite{foucart2013mathematical}.

However, in order to develop an expression to quantify the convergence criterion, we can leverage the sparsity in the rows of our sensing matrix \(\bm{A}^{\left( u,m \right)}\ \), which is due the limited number of RF chains. To this end, by neglecting the phase of the pilot symbols in the sensing matrix, we can focus on the convergence criterion for a pessimistic case where information recovery depends only on the non-zeros elements in each column (i.e., the random beam selections).

We can then claim that the RIP is met with submatrix conditioning owing to at least one pilot symbol separation between columns, if the following are met:

\begin{enumerate}
	\def\labelenumi{\arabic{enumi})}
	\item
	There is at least a single non-zero value in each column of
	\(\bm{A}^{\left( u,m \right)}\ \).
	\item
	No two columns of \(\bm{A}^{\left( u,m \right)}\ \)are identical.
\end{enumerate}
For the first of these conditions, recall that FPA was already proposed to reweight the candidate beam selection probabilities and ensure that \(\bm{A}^{\left( u,m \right)}\) has at least one non-zero value in each column. As such, the first condition is always met before attempting estimation with GAMP, and therefore can always be guaranteed in our results.

Turning to the second condition, consider that in each length-\(N_{\text{BS}}N_{\text{UE}}\) row of the sensing matrix, there are \(R_{\text{BS}}\) non-zero elements that are randomly selected. As the FPA approach works to spread the non-zero elements uniformly across each column, we can approximate the probability that any column has a non-zero entry in any given row as
\begin{align}
\Pr\left( \ \left| \bm{A}_{r,c}^{\left( u,m \right)} \right| > 0\  \right) \cong \frac{R_{\text{BS}}}{N_{\text{BS}}N_{\text{UE}}}\, \ \forall\ c = 1,\ldots N_{\text{BS}}N_{\text{UE}}
\end{align}
Then for any arbitrary number of measurements \(m = T_{E}R_{\text{UE}}\), we can express the expectation of the number of non-zero entries in each column as
\begin{align}
E_{\text{NZ}} = \sum_{r = 1}^{T_{E}R_{\text{UE}}}{\Pr\left( \ \left| \bm{A}_{r,c}^{\left( u,m \right)} \right| > 0\  \right)}
= \frac{R_{\text{BS}}R_{\text{UE}}}{N_{\text{BS}}N_{\text{UE}}}T_{E} ,
\end{align}
$\forall\ c = 1,\ldots N_{\text{BS}}N_{\text{UE}}$.
For each column with length-\(T_{E}R_{\text{UE}}\ \)and with \(E_{\text{NZ}}\) non-zero entries, the number of uniquely possible combinations is given by
\begin{align}
\ C_{\text{NZ}} = \frac{\left( T_{E}R_{\text{UE}} \right)!}{E_{\text{NZ}}!\left( T_{E}R_{\text{UE}} - E_{\text{NZ}} \right)!}.
\end{align}
Relating this back to the convergence, we can now find the probability all columns in \(\bm{A}^{\left( u,m \right)}\) are unique.

Although FPA will inherently reweight the probabilities to reduce the chance of identical columns, we take a pessimistic approach and consider the distribution of non-zero elements in each column to be independent. With this simplification, the probability that all \(N_{\text{BS}}N_{\text{UE}}\ \) of the columns in \(\bm{A}^{\left( u,m \right)}\) are selected uniquely can be
expressed by
\begin{align}
P_{U} &= \frac{C_{\text{NZ}}\ }{C_{\text{NZ}}\ } \times \ \ldots \times \frac{C_{\text{NZ}} - \left( N_{\text{BS}}N_{\text{UE}} - 1 \right)}{C_{\text{NZ}}\ } \nonumber \\
&= \prod_{n = 0}^{N_{\text{BS}}N_{\text{UE}} - 1}\left( 1 - \frac{{n\ }}{C_{\text{NZ}}\ } \right)
\end{align}
By substituting \(C_{\text{NZ}}\) and \(E_{\text{NZ}}\) and relaxing the
factorials to gamma functions, $P_U$ then provides an upper bound for the
probability of convergence after \(T_{E}\) time slots, denoted by $P_c$, as (\ref{converg}) on top of the next page.
\begin{figure*}
\begin{align} \label{converg}
P_{U} = \prod_{n = 1}^{N_{\text{BS}}N_{\text{UE}} - 1}\left( 1 - n\ \frac{\ \Gamma\left( \frac{T_{E}R_{\text{UE}}R_{\text{BS}}}{N_{\text{UE}}N_{\text{BS}}} + 1 \right) \Gamma\left( T_{E}R_{\text{UE}}\left( 1 - \frac{R_{\text{BS}}}{N_{\text{UE}}N_{\text{BS}}} \right) + 1 \right)}{\Gamma\left( T_{E}R_{\text{UE}} + 1 \right)} \right)\ge P_c.
\end{align}
\hrulefill
\vspace*{4pt}
\end{figure*}

It is worth noting that although this analysis has not directly considered the proposed PEPA strategy, because PEPA is only ever applied after FPA has forced the convergence conditions and it also follows the same upper bound.

\color{black}

\section{Numerical Results}


We now provide some numerical results to evaluate the performance of our proposed SWIFT algorithm. We consider a mmWave system with $N_{\!B\!S} = 32$ antennas at the BS and $N_{\!U\!E} = 16$ antennas at each user. We further consider the BS to be equipped with $R_{\!B\!S}=8$ RF chains and each user to be equipped with $R_{\!U\!E}=4$ RF chains. We consider the expected number of paths to be $E[L^{(u)}]=3$ with AOD and AOD uniformly distributed on the continuous range $[0,2\pi]$. We also set the maximum allowed number of measurements the same as the exhaustive search-based approach, i.e., $T_{max}=N_{\!B\!S}N_{\!U\!E}/R_{\!U\!E}$. We update the channel estimate every $T_u=N_{\!U\!E}/R_{\!U\!E}=4$ measurements and use $\Gamma=10^{-1}$ in the binarization process of the estimated channel vector. We show two variants of the SWIFT algorithm. The first uses the forcing probability adaptation approach proposed in Sec. \ref{ssec:sec_forcing} and is labeled as SWIFT-FPA, whereas the second variant uses the partially estimated probability adaptation from Sec. \ref{ssec:sec_pepa} and is labeled as SWIFT-PEPA.


\begin{figure}[!t]
	\centering
	\includegraphics[width=3.3in,trim={1.2cm 6.9cm 1.5cm 9.0cm},clip]{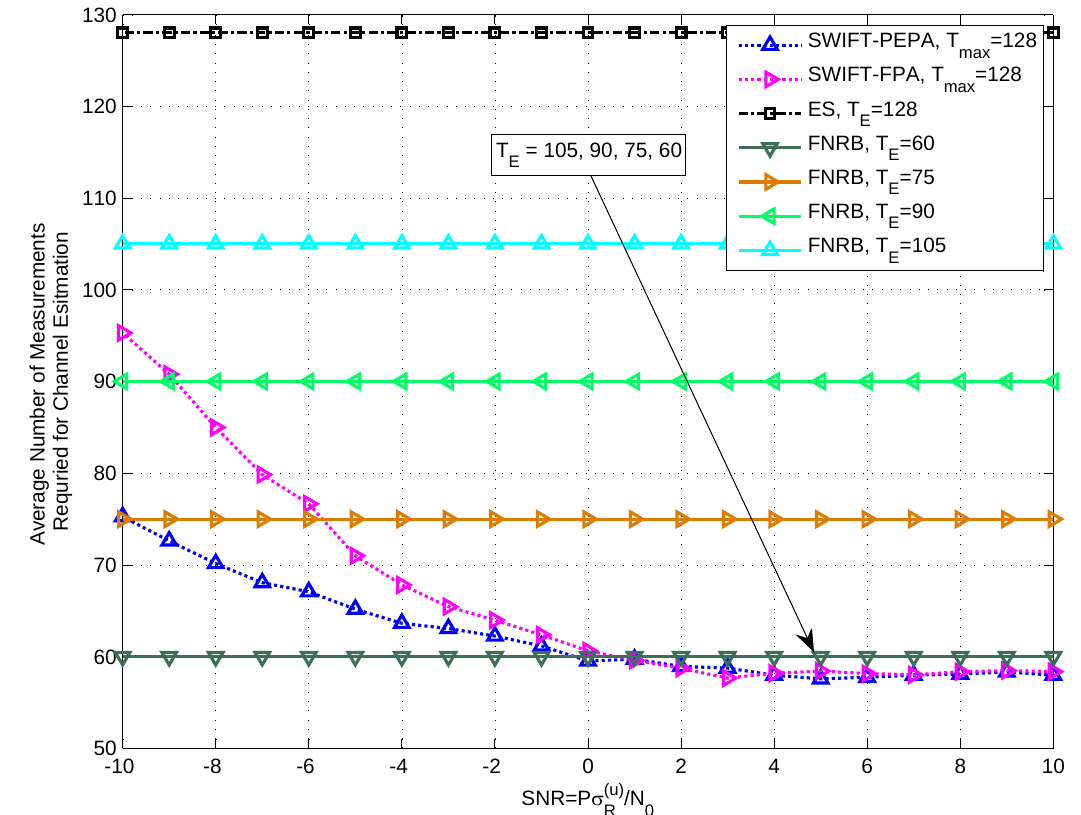}
	\caption{Average number of measurements required for channel estimation when the BS is equipped with $N_{\!B\!S}=32$ antenna and $R_{\!B\!S}=8$ RF chains and the user is equipped with $N_{\!U\!E}=16$ antenna and $R_{\!U\!E}=4$ RF chains. We assume the number of paths is $E[L^{(u)}]=3$ and update the channel estimate every $T_u=4$ measurements.}
	\label{SU_M}
\end{figure}

Recall that single-user oriented angular refinement approaches such as \cite{rheath,7579573} are no longer suitable for simultaneous multi-user estimation. In order to compare our proposed algorithm to a simultaneous estimation scheme, we consider a random beamforming-based channel estimation approach that uses a predetermined number of measurements. As these fixed-number random beamforming (FNRB) approaches do not target any specific user, the training overhead does not scale as the number of users is increased \cite{alkhateeby2015compressed}. The adopted GAMP estimator used in SWIFT is also applied in the FNRB schemes to estimate the channel information. We also compare our scheme with a benchmark exhaustive search-based approach, in which an estimate of the virtual channel can be found by individually measuring the gains between all combinations of the candidate vectors (i.e., transmitting with only a single beamforming vector but receiving with $R_{\!U\!E}$ beamforming vector(s) in each measurement). We represent this approach by exhaustive search (ES) in all figures.

We first show simulation results for the single-user case in Fig. \ref{SU_M} over a range of different SNR values. It is worth noting that the single-user case is the equivalent to the multi-user case where the BS can communicate with each user in an adjacent dedicated sub-channel, as discussed in Sec. \ref{sec:BS_stop}. Fig. \ref{SU_M} shows the average number of channel measurements by each of the aforementioned approaches. We can see that both variations of the SWIFT algorithm are able to adaptively increase the number of measurements at low SNR values in order to meet the required channel estimation convergence criterion. As all other algorithms use a fixed number of measurements, their average number of measurements remains unchanged across the whole SNR range. Comparing the two SWIFT approaches, we see that SWIFT-PEPA requires less measurements at low SNR. We also see that the two schemes converge to a similar average number of channel measurements at high SNRs.

\begin{figure*}[!t]
\centering
\subfigure[][]{\includegraphics[width=3.1in,trim={1.2cm 6.9cm 1.5cm 8.0cm},clip]{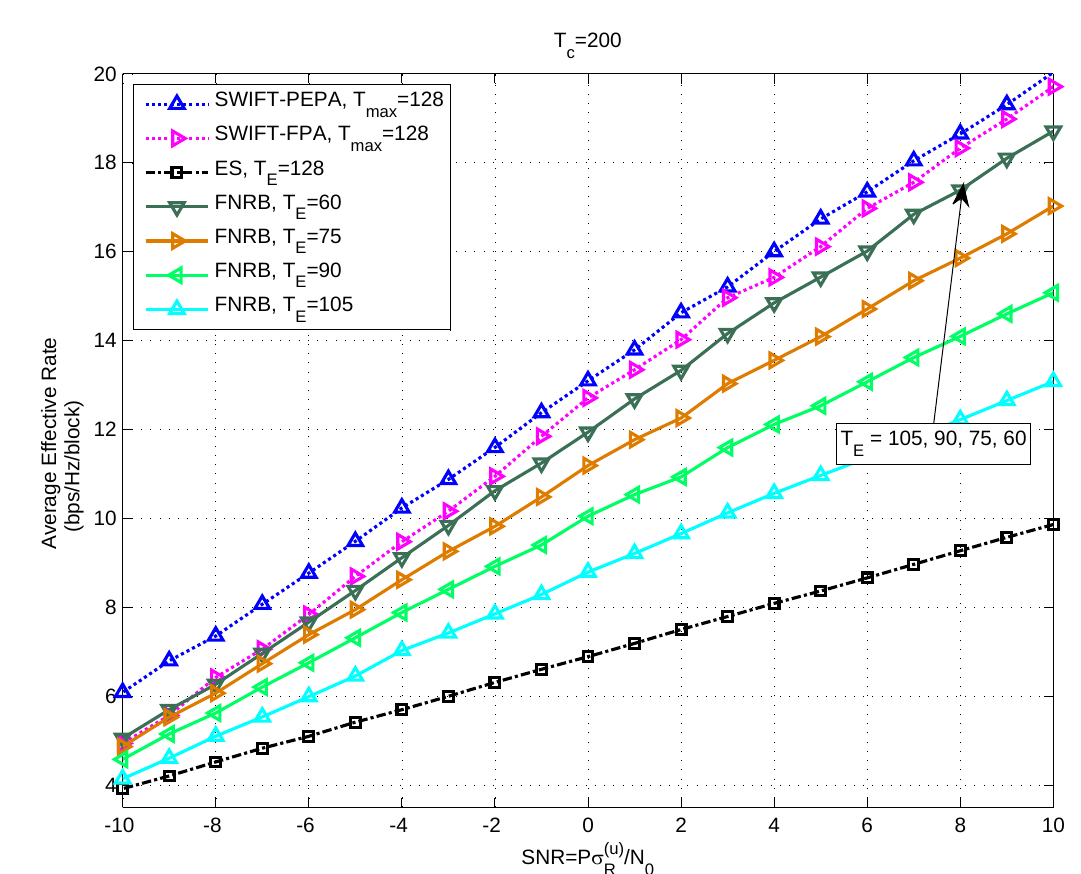}}
\subfigure[][]{\includegraphics[width=3.1in,trim={1.2cm 6.9cm 1.5cm 8.0cm},clip]{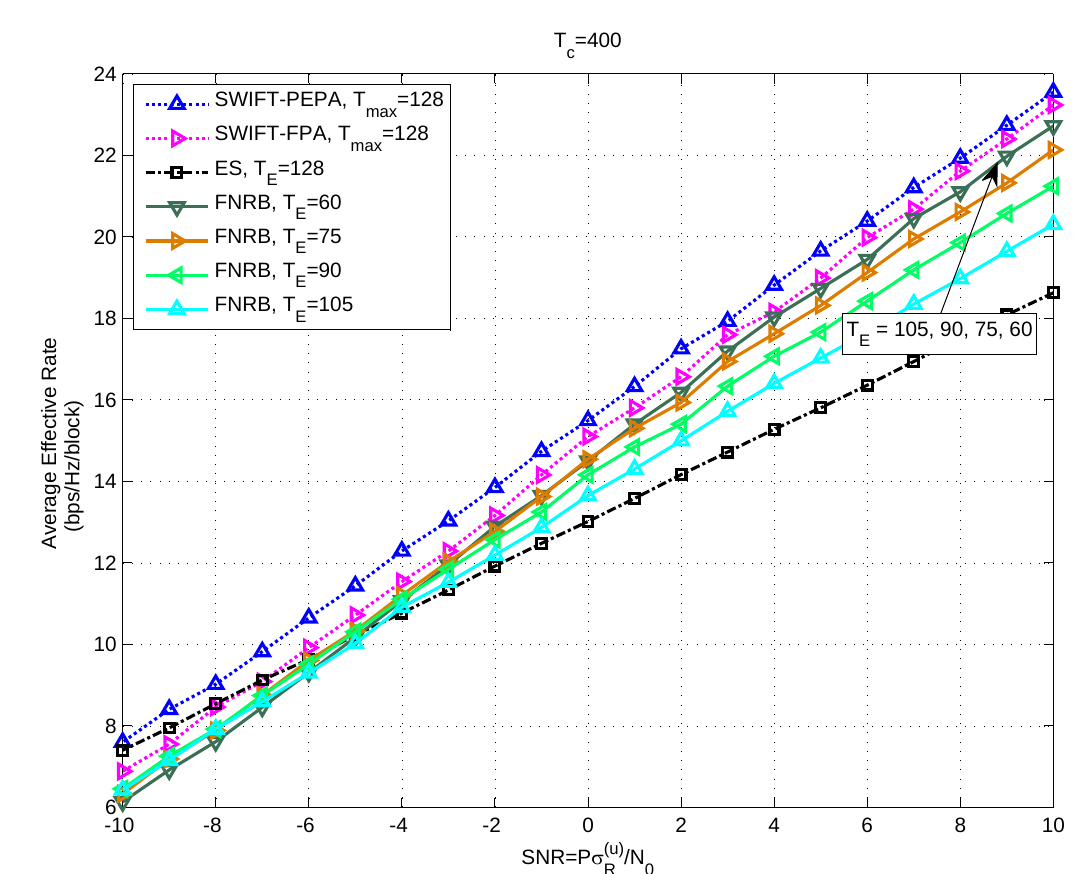}}
\caption{Single-user effective rate for (a) $T_c=200$ and (b) $T_c=400$ when the BS is equipped with $N_{\!B\!S}=32$ antennas and $R_{\!B\!S}=8$ RF chains and the user is equipped with $N_{\!U\!E}=16$ antennas and $R_{\!U\!E}=4$ RF chains. We assume the number of paths is $\text{E}[L^{(u)}]=3$ and update the channel estimate every $T_u=4$ measurement time slots.}
\label{SU_rate}
\end{figure*}

Fig. \ref{SU_rate} shows the resulting average effective rate as defined by (\ref{effective_rate}) of various schemes with different length of coherence time\footnote{\edit{Although MSE is a common estimation performance metric, for channels with both strong and even imperfect sparsity, it can often favor conservative algorithms that yield weak and incorrect results over ones that actually attempt to provide a solution.}}. More specifically, Fig. \ref{SU_rate} (a) considers a coherence time of $T_c =200$ and Fig. \ref{SU_rate} (b) considers $T_c =400$ symbols. As the adopted performance metric of effective rate considers both the training quality and overhead, it more accurately reflects the channel estimation performance. From Fig. \ref{SU_rate} (a), we can observe that both SWIFT approaches are able to achieve a superior effective rate over a large range of SNR values. We also see that different FNRB schemes using a fixed number of measurements can outperform each other depending on both the value of SNR and coherence time. In particular, this can be seen in Fig. \ref{SU_rate} (b) where FNRB with $T_E=60$ is the best performing scheme at high SNR, but the worst at low SNR. In contrast, SWIFT-PEPA is always the best performing scheme. It is worth noting that the complexity of SWIFT-PEPA is slightly higher than SWIFT-FPA, as the beam selection probability vector is based on the channel measurements and therefore cannot be computed offline.

\begin{figure}[!t]
\centering
\includegraphics[width=3.3in,trim={1.2cm 6.9cm 1.5cm 9.0cm},clip]{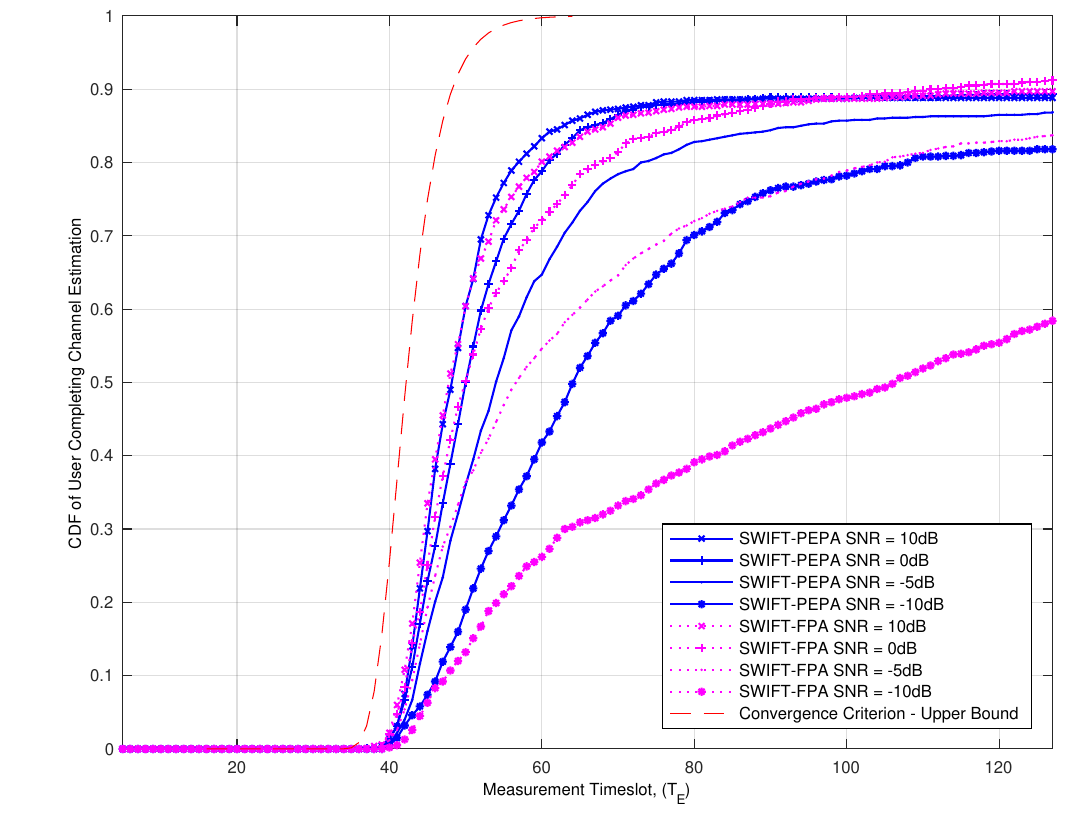}
\caption{Cumulative distribution functions for users with different SNR completing their channel estimation before $T_E$.}
\label{P_CDF}
\end{figure}

In order to gain an insight into when a user is likely to complete its channel estimation, we plot in Fig. \ref{P_CDF} the cumulative distribution function (CDF) that a user completes its channel estimation before a given duration $T_E$ for both SWIFT approaches with various SNR values. From Fig. \ref{P_CDF}, we can first see that users at a larger SNR are more likely to estimate their channel before those with low SNR. With a large number of users distributed across all SNRs, we can infer that the occurrence of channel estimation feedback events would be spread over a large number of different times. As a result, this would alleviate pressure on the feedback channel (used by each user to feedback beamforming directions to the BS) as the number of users needing to communicate at any given time would be significantly reduced. It is also interesting to compare the differences in CDFs for both SWIFT variations. It can be seen that at high SNR, both approaches have similar CDFs which is consistent with the observation in Fig. \ref{SU_M} where both approaches are seen to have the same average number of measurements. At low-to-medium SNR, it can be seen that, by adopting SWIFT-PEPA, users having a greater probability of completing their estimation with a shorter duration. This is again consistent with Fig. \ref{SU_M}, where SWIFT-PEPA has a lower average number of measurements in the low-to-medium SNR ranges. \edit{To validate our convergence analysis,  Fig. \ref{P_CDF} also shows the upper bound for the probability of convergence in (\ref{converg}). As can seen, both SWIFT-FPA and SWIFT-PEPA satisfy the bound. At larger values of \(T_{E}\ \), the numerical CDFs do not approach the convergence bound due to the non-zero probability of the channel either being in a deep fade or having no paths, which were not considered in this convergence bound.}



We now turn to a multi-user scenario with $U$ users in a single cell of radius $R$. We assume that the $u$th user has a distance $d^{(u)}$ from the BS and this distance is uniformly distributed within the range $[0,R]$. We then model the variance of the fading coefficient for the $u$th user as a function of distance by $\sigma_R^{(u)} = (d^{(u)})^{-\beta}$, where $\beta$ is the path loss exponent. As the distances between BS and users are not expected to change rapidly relative to the cell size, we consider that $\sigma_R^{(u)}$ is known to each user from experience of previous channels\footnote{It is worth pointing out that the channel estimation tools used in this paper can also be extended to jointly estimate the channel statistics in a similar way as in \cite{vila2011expectation}.}. We set the BS transmit power $P=20$dBm, the noise power $N_0=-60$dBm, the path loss exponent $\beta=4$ and the cell radius $R=200$m. For example, this configuration leads to $P\sigma_R^{(u)}/N_0=-12$dB at $d=200$m and $P\sigma_R^{(u)}/N_0=12$dB at $d=50$m.

We evaluate the multi-user performance when the channel estimation and data communication share the same frequency, i.e., all BS pilots must stop in order for communication to commence. To this end, we consider that the BS is only able to communicate with $N_{s}=10$ users in a given coherence block. Note that the other schemes only know the beamforming directions, they randomly select $N_{s}=10$ users from those who send back the beamforming directions. In contrast, the SWIFT approaches wait until the first $N_{s}=10$ users have fed back requests for communication and then the BS begins communicating with them. For simplicity, in all schemes, we consider that the BS divides the remaining communication time equally among the users.

\begin{figure*}[!t]
\centering
\subfigure[][]{\includegraphics[width=3.1in,trim={1.2cm 6.9cm 1.5cm 8.0cm},clip]{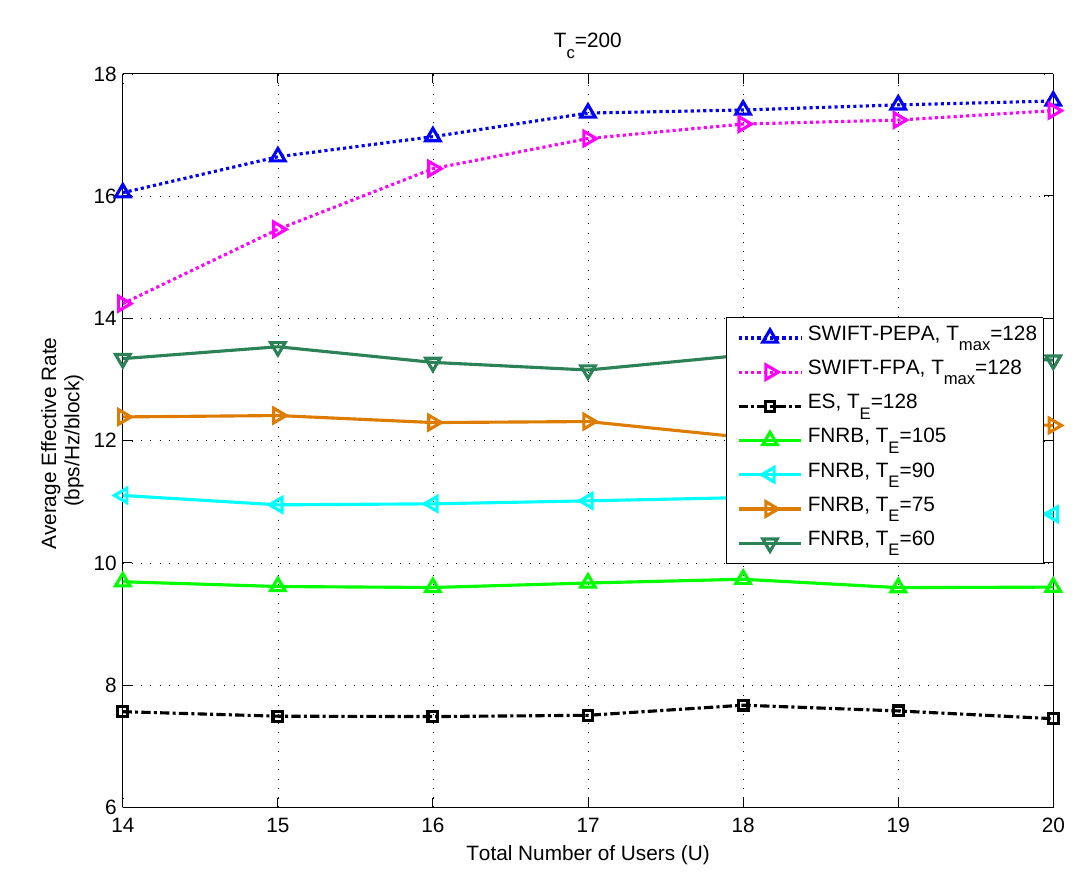}}
\subfigure[][]{\includegraphics[width=3.1in,trim={1.2cm 6.9cm 1.5cm 8.0cm},clip]{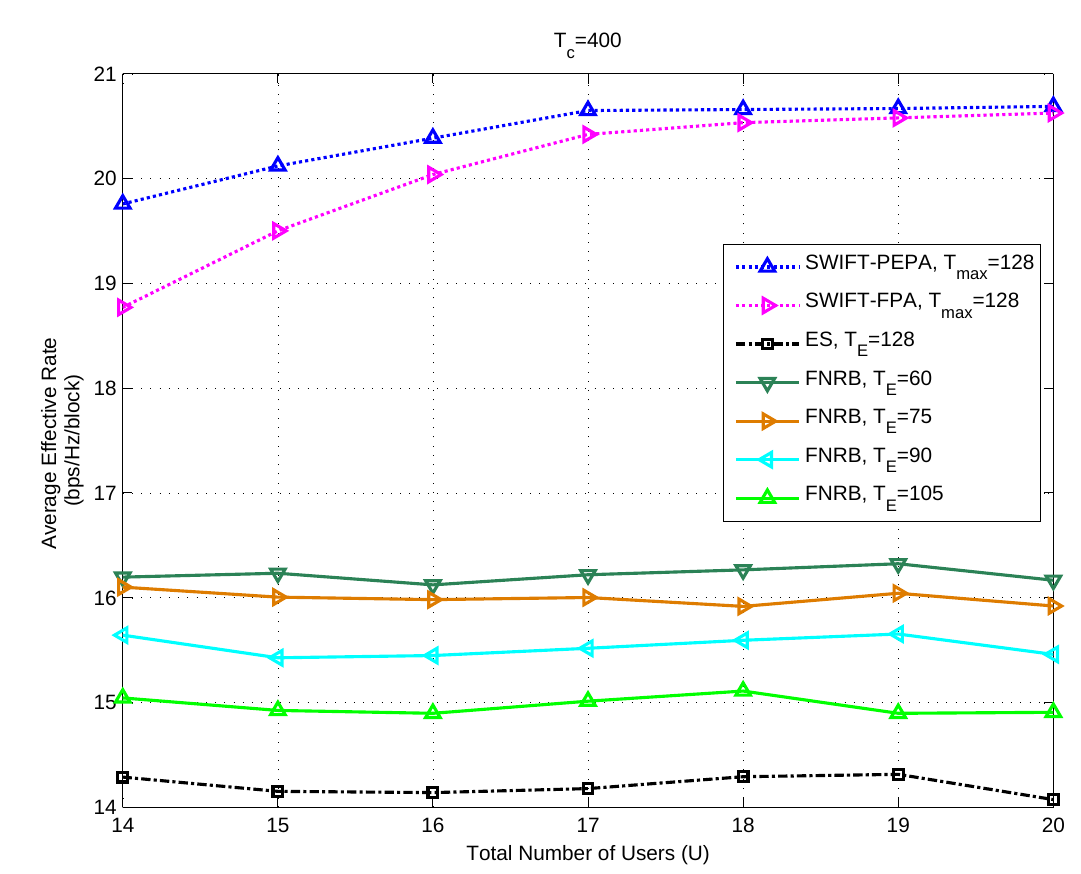}}
\caption{Average per-user effective rate for multi-user scenario in Fig. \ref{deployment_scenario} with (a) $T_c=200$ and (b) $T_c=400$ when the BS is equipped with $N_{\!B\!S}=32$ antennas and $R_{\!B\!S}=8$ RF chains and the user is equipped with $N_{\!U\!E}=16$ antennas and $R_{\!U\!E}=4$ RF chains. We assume the number of paths is $\text{E}[L^{(u)}]=3$ and update the channel estimate every $T_u=4$ measurement time slots.}
\label{MU_rate}
\end{figure*}

Fig. \ref{MU_rate} shows the average effective rate as the number of users increase for two scenarios with (a) $T_c=200$ and (b) $T_c=400$. We first see that the average effective rate increases significantly as the number of users increases. This is because the SWIFT algorithm is capable of selecting out users with channels better suited for data communication based on the sequence that users feedback their channel estimation. In contrast, the other schemes remain unchanged as the number of users increases. Furthermore, the effective rate of the SWIFT schemes begins to saturate from about $U=17$. That is, as the BS only communicates with the first $N_{s}=10$ users, it only needs to neglect a little more than one third of these users to achieve a significant performance gain. Given the system model, these neglected users may be near the cell edge and may have a more favorable channel with an adjacent cell. \edit{However, in other cases, to more generally provide fairness among all UE, a more advanced scheduling approach may be required to favor those with poorer channels. As our scheme has focused on improving the overall network performance as opposed to individual UE, this has been left as a future work.}



\section{Conclusions}
In this paper we have proposed a novel Simultaneous-estimation With Iterative Fountain Training (SWIFT) framework for multi-user channel estimation in mmWave MIMO communication systems. In the proposed algorithm, additional measurements are carried out in an adaptive manner when required, allowing the channel estimate to converge to the predetermined accuracy. We have shown that the proposed approach yields superior effective rate performance when compared to those random beamforming-based approaches with fixed number of measurements. By utilizing the users' order in terms of completing their channel estimation, we have also shown our SWIFT framework can infer the sequence of users' channel quality and perform the associated user scheduling to achieve superior performance, especially for resource-constrained scenarios where only a limited number of users can be served.

\linespread{1}
\bibliographystyle{IEEEtran}
\bibliography{IEEEabrv,\jobname}


\vspace{-1.2cm}

\end{document}